\documentclass[aps,prx,twocolumn,amsmath,amssymb,nofootinbib,superscriptaddress,floatfix,reprint,longbibliography]{revtex4-1}
\usepackage[dvips]{graphicx}
\usepackage{latexsym}
\usepackage{amsmath}
\usepackage{amsfonts}
\usepackage{amssymb}
\usepackage{bm}
\usepackage{color}
\usepackage{txfonts}
\usepackage{float}
\usepackage{url}
\usepackage{diagbox}
\usepackage[colorlinks=true, urlcolor=blue, linkcolor=blue, citecolor=blue, pdftex]{hyperref}
\usepackage{ulem}
\usepackage{physics}
\usepackage{pifont}
\begin{document}
	\newcommand{\fig}[2]{\includegraphics[width=#1]{#2}}
	\newcommand{\pprl}{Phys. Rev. Lett. \ }
	\newcommand{\pprb}{Phys. Rev. {B}}

\title {Magnetic electron-hole asymmetry in cuprates: a computational revisit}

\author{Jiong Mei}
\thanks{These authors contributed equally.}
\affiliation{Beijing National Laboratory for Condensed Matter Physics and Institute of Physics,
	Chinese Academy of Sciences, Beijing 100190, China}
\affiliation{School of Physical Sciences, University of Chinese Academy of Sciences, Beijing 100190, China}

\author{Shao-Hang Shi}
\thanks{These authors contributed equally.}
\affiliation{Beijing National Laboratory for Condensed Matter Physics and Institute of Physics,
	Chinese Academy of Sciences, Beijing 100190, China}
\affiliation{School of Physical Sciences, University of Chinese Academy of Sciences, Beijing 100190, China}

\author{Ping Xu}
\thanks{These authors contributed equally.}
\affiliation{Key Laboratory of Artificial Structures and Quantum Control (Ministry of Education), School of Physics and Astronomy, Shanghai Jiao Tong University, Shanghai 200240, China}

\author{Ziyan Chen}
\thanks{These authors contributed equally.}
\affiliation{Beijing National Laboratory for Condensed Matter Physics and Institute of Physics,
	Chinese Academy of Sciences, Beijing 100190, China}
\affiliation{School of Physical Sciences, University of Chinese Academy of Sciences, Beijing 100190, China}

\author{Hui-Ke Jin}
\email{jinhk@shanghaitech.edu.cn}
\affiliation{School of Physical Science and Technology, ShanghaiTech University, Shanghai 201210, China}

\author{Mingpu Qin}
\email{qinmingpu@sjtu.edu.cn}
\affiliation{Key Laboratory of Artificial Structures and Quantum Control (Ministry of Education), School of Physics and Astronomy, Shanghai Jiao Tong University, Shanghai 200240, China}
\affiliation{Hefei National Laboratory, Hefei 230088, China}

\author{Zi-Xiang Li}
\email{zixiangli@iphy.ac.cn}
\affiliation{Beijing National Laboratory for Condensed Matter Physics and Institute of Physics,
	Chinese Academy of Sciences, Beijing 100190, China}
\affiliation{School of Physical Sciences, University of Chinese Academy of Sciences, Beijing 100190, China}

\author{Kun Jiang}
\email{jiangkun@iphy.ac.cn}
\affiliation{Beijing National Laboratory for Condensed Matter Physics and Institute of Physics,
	Chinese Academy of Sciences, Beijing 100190, China}
\affiliation{School of Physical Sciences, University of Chinese Academy of Sciences, Beijing 100190, China}

\date{\today}

\begin{abstract}
In this work, we revisit the electron–hole asymmetry of antiferromagnetism in cuprates by studying the three-band Emery model. Using parameters relevant to La$_2$CuO$_4$, we benchmark the anti-ferromagnetic response for a large range of dopings with variational Monte Carlo, determinant quantum Monte Carlo, constrained-path auxiliary-field quantum Monte Carlo, density-matrix embedding theory, and the Gutzwiller approximation. Across methods and accessible sizes/temperatures, we find no significant electron–hole asymmetry if we consider only Neel anti-ferronagnetic response and ignore other possible orders such as stripe state. This result is robust to a moderate oxygen-site repulsion $U_p$ and to parameter sets of Nd$_2$CuO$_4$. Incorporating dopant-induced local potentials reveals an extrinsic route to asymmetry: Cu-site defects enhance AFM on the electron-doped side, whereas O-site defects suppress it on the hole-doped side. These results indicate that dopant-driven effects make a non-negligible contribution to apparent electron–hole asymmetry in the general phase diagram of cuprates and should be included when analyzing competing orders in cuprates.

\end{abstract}
\maketitle

\section{Introduction}
\begin{figure}[t]
	\begin{center}
		\fig{3.4in}{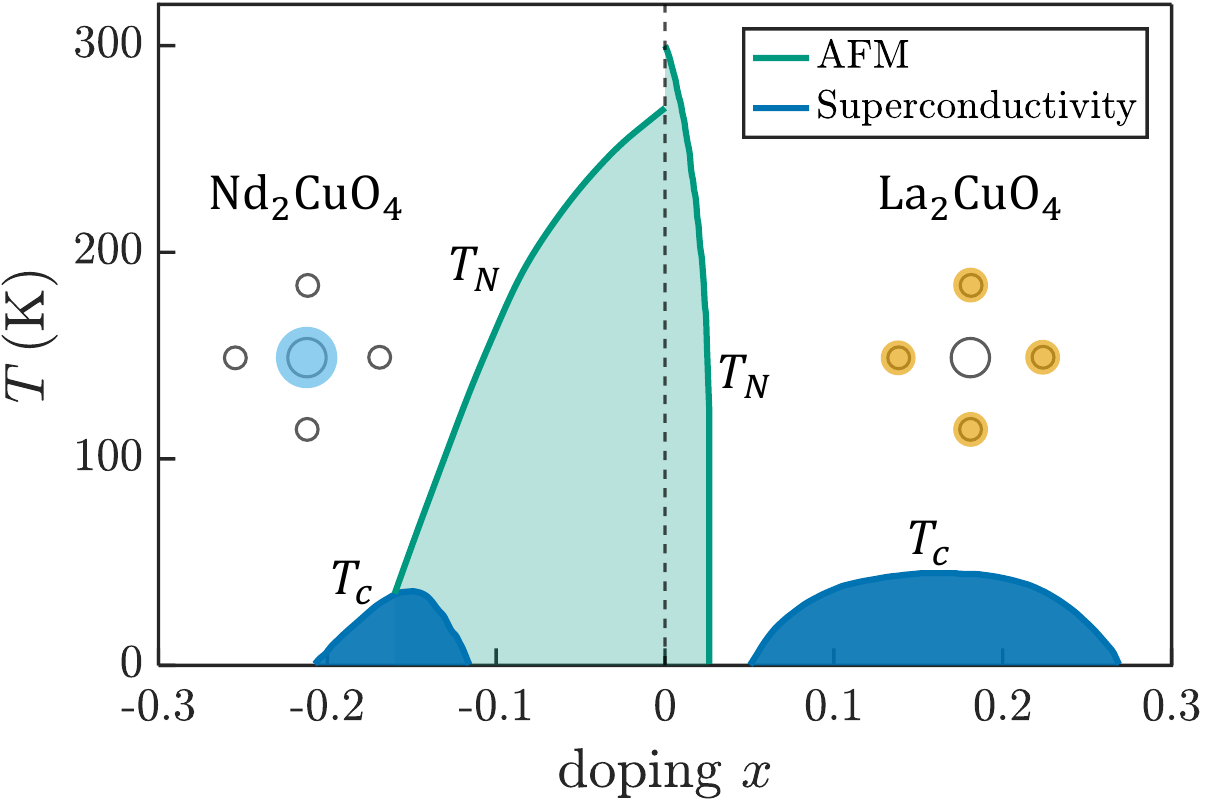}
		\caption{ Schematic phase diagram of hole-doped and electron-doped cuprates. Only the antiferromagnetism ($T_N$) and superconductivity phase boundaries ($T_c$) are plotted. Experimental data is summarized in Ref. \cite{armitage_RevModPhys.82.2421}. Right: hole-doped La$_2$CuO$_4$; Left: electron-doped Nd$_2$CuO$_4$.
        It is noted that both maximum $T_N$ and maximum $T_c$ are comparable in these two materials. But the ending point of AFM phase varies strongly, leading to the electron-hole asymmetry. In the electron-doped side, the AFM ending point is still unclear \cite{armitage_RevModPhys.82.2421}. In cuprates, doped holes tend to reside on the oxygen sites (the smaller circle in the right), while doped electrons are mainly located on the copper sites (the larger circle in the left).
			\label{fig1}}
	\end{center}
\end{figure}

The emergence of high-temperature superconductivity in cuprates presents one of the central mysteries in modern condensed matter physics \cite{bednorz,doping_mott,keimer_review,Anderson_RVB,armitage_RevModPhys.82.2421,scalapino_RevModPhys.84.1383,Norman_2003}. Unlike conventional superconductors, the unconventional superconductivity arises from doping an antiferromagnetic Mott insulator, where strong electron-electron correlations play a crucial role \cite{doping_mott,Anderson_RVB,Zhang_Rice,keimer_review}. 
As a result, understanding the impact of these correlations remains a major theoretical and experimental difficulty. Cuprates have thus become the standard system for studying strongly electron correlation. 
One of their most intriguing features is that, despite differences in chemical composition and crystal structure, various cuprate compounds display remarkably similar phase diagrams and electronic properties \cite{armitage_RevModPhys.82.2421}, as illustrated in Fig.~\ref{fig1}.

Cuprate superconductors are broadly classified into two types based on the dopants \cite{armitage_RevModPhys.82.2421,Xiang_Wu_2022}: hole-doped and electron-doped systems. 
Here, we take La$_2$CuO$_4$ (LCO) \cite{bednorz,lco_PhysRevB.45.7430} as a representative hole-doped compound and Nd$_2$CuO$_4$ (NCO) \cite{nco_PhysRevLett.62.1197,nco_PhysRevLett.93.027002} as a representative electron-doped compound in Fig.~\ref{fig1}.
Interestingly, despite having comparable maximum Neel temperatures $T_N$ (320 K for LCO \cite{lco_PhysRevB.45.7430}, 270 K for NCO \cite{nco_PhysRevLett.93.027002}) and maximum superconducting transition temperatures $T_c$ (36 K for doped-LCO \cite{lco_tc_PhysRevLett.61.1127} , 24 K for doped-NCO \cite{nco_PhysRevLett.62.1197}), their phase diagrams reveal a pronounced electron-hole asymmetry in Fig.~\ref{fig1}. In particular, the antiferromagnetic (AFM) order in La$_2$CuO$_4$ is suppressed rapidly and disappears around 3\% hole doping, whereas in Nd$_2$CuO$_4$, AFM persists up to approximately 15\% electron doping \cite{armitage_RevModPhys.82.2421,kastner_RevModPhys.70.897}. The origin of this electron-hole asymmetry remains an open and debated question in the study of cuprates.

One essential difference between electron and hole doping comes from the charge-transfer nature of cuprates \cite{Zannen_PhysRevLett.55.418,Emery_PhysRevLett.58.2794,Zhang_Rice}. 
One key energy scale for cuprates is the charge-transfer energy between $d$ electrons and their neighbor $p$ electrons $\Delta_{CT}$.
As a result of the charge-transfer nature of cuprates, \textit {the doped holes in cuprates reside primarily on O sites while the doped electrons reside primarily on Cu sites}.
This charge-transfer effect can be well captured by the three-band Emery model \cite{Emery_PhysRevLett.58.2794}. A single-band low energy effective $t-J$ (or Hubbard) model can be derived from the three-band model, owing to Zhang-Rice's work \cite{Zhang_Rice}. In the single band model, the presence of the second nearest neighbor hopping $t'$ results in the electron and hole doping asymmetry \cite{Maekawa_JPSJ.59.1760,Hybertsen_PhysRevB.41.11068,Feiner_PhysRevB.45.7959,Belinicher_PhysRevB.50.13768,Dagotto_RevModPhys.66.763,tj_PhysRevB.49.3596,tj_PhysRevB.50.12866,tj_PhysRevB.64.212505,tj_PhysRevB.65.134414,tj_PhysRevLett.102.027002} and the electron doping and hole doping cases can be mapped to each other by reversing the sign of $t'$ through a particle-hole transformation.

In this work, we revisit the long-standing problem of electron–hole asymmetry in cuprates by accurate quantum many-body computation. We directly study the three-band model and calculate both the electron and hole doping cases. Recent advances in quantum many-body techniques have enabled detailed studies of correlated models such as the $t$–$J$ and single-band Hubbard models \cite{White1999PRB,Corboz2014PRL,Devereaux2016PRB,Zheng2017Science,qin_PhysRevX.10.031016,t-j-dmrg,hubbard_review,Huang2018npjQM,Jiang2019Science,Huang2019Science,PhysRevB.102.041106,Jiang2020PRR,Gong2021PRL,He2021PRX,Li2024PRL,He2023PRX,White2023PRB,Gong2024PRL,Jiang2024arXiv,Xu2024Science,Shen_2025,Liu2025PRL,Delft2025PRL,Yao2025SB,Sheng2025PNAS,Lv2025arXiv,Gu2025arXivtjmodel,Devereaux2025arXiv}. We benchmark and cross-validate different approaches \cite{PhysRevX.5.041041} in the three band Hamiltonian including Variational Monte Carlo (VMC) \cite{vmc_PhysRevB.16.3081,vmc_shiba}, Determinant Quantum Monte Carlo (DQMC) \cite{DQMC_PhysRevD.24.2278,DQMC_PhysRevLett.46.519,DQMC_PhysRevLett.47.1628}, Constrained-Path Auxiliary-Field Quantum Monte Carlo (CP-AFQMC) \cite{AFQMC_PhysRevB.55.7464,AFQMC_PhysRevB.55.7464}, Density Matrix Embedding Theory (DMET) \cite{DMET_PhysRevLett.109.186404,DMET_guide}, and the Gutzwiller Approximation (GA) \cite{GW_PhysRev.137.A1726,GW_PhysRevLett.10.159,KR_PhysRevLett.57.1362}—to investigate the magnetic properties of cuprates. It is well established that intertwined orders are an intrinsic aspect of cuprate physics \cite{intertwined,Fradkin_RevModPhys.87.457}, which poses significant challenges for numerical simulations. Among these competing phases, however, Néel order is the most accessible symmetry-broken state to capture. To simplify the discussion and highlight the magnetic electron–hole asymmetry, we therefore focus exclusively on the AFM solution in this work, deferring the study of other competing orders \cite{YANAGISAWA2002292,Aoki1996PRL,White_PhysRevB.92.205112,PhysRevB.102.214512,PhysRevB.108.205154,Huang2017Science,doi:10.1073/pnas.2408717121,Varma2014PRL} to future investigations. We also analyze the effect of dopant-induced local potential on the AFM order.


This paper is organized as follows. In Section~\ref{sec:model}, we introduce the three-band model and specify the parameters used in our calculations. Section~\ref{sec:methods} provides a brief overview of the computational methods employed, with particular emphasis on how magnetic properties are extracted within each approach. Our main results on the electron–hole magnetic phase diagram, along with its dependence on model parameters, are presented in Section~\ref{sec:phase-diagram}. The influence of the doping potential on the phase diagram is discussed in Section~\ref{sec:doping}. Finally, we summarize our findings and outline future directions in Section~\ref{sec:summary}. Additional technical details and supplementary results are provided in the Appendices.

\section{Three-band model and parameters}
\label{sec:model}

\begin{figure}[t]
  \centering
  \includegraphics[width=0.48\textwidth]{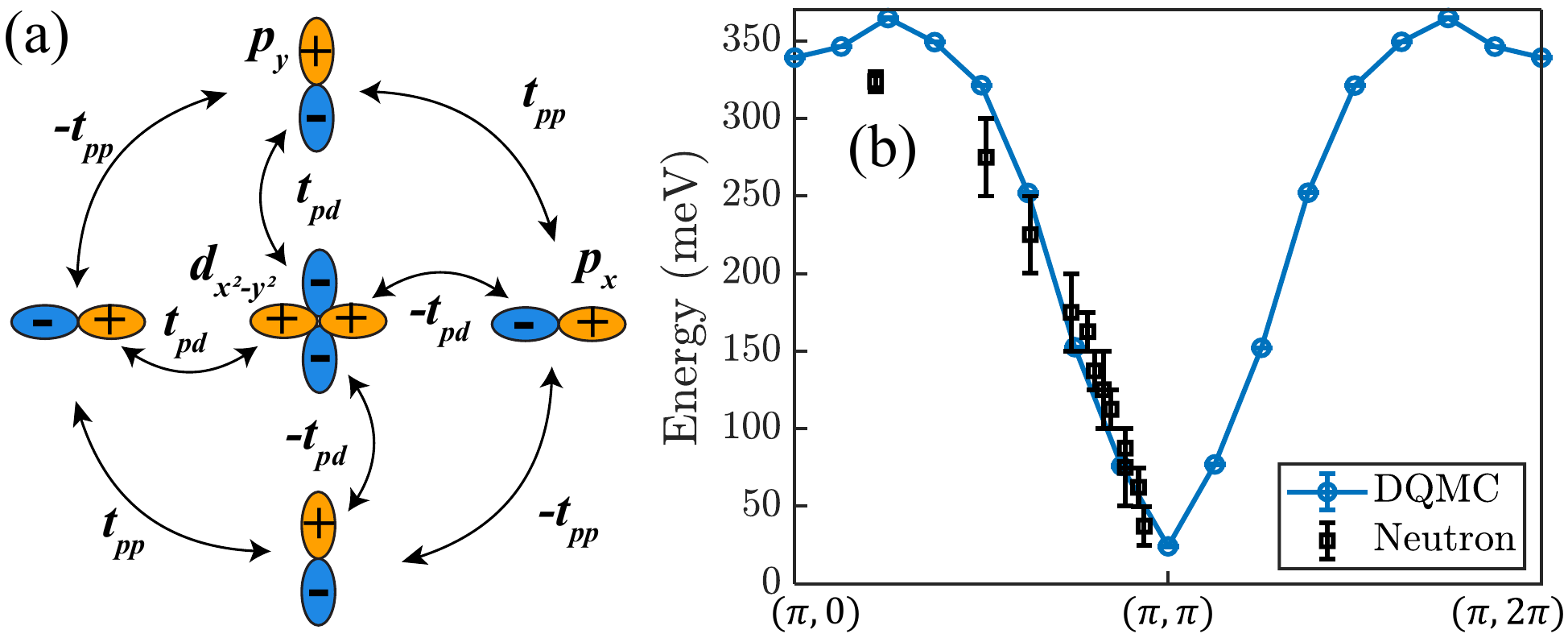}
  \caption{(a) Schematic of the three-band Emery model on the CuO$_2$ plane in the hole representation. Orange/blue lobes indicate the orbital phase of Cu $d_{x^2-y^2}$ and O $p_{x/y}$. The signs of $t_{pd}$ and $t_{pp}$ follow the indicated phases. (b) Spin-wave dispersion obtained from DQMC on a $16\times4$ lattice at inverse temperature $\beta=8$. Blue circles denote the DQMC results, and black squares correspond to the experimental dispersion measured by inelastic neutron scattering in Ref.~\cite{La2CuO4_PhysRevLett.86.5377}.}
  \label{fig2}
\end{figure}

In cuprates, the essential structural and electronic component is the CuO$_2$ plane, as illustrated in Fig.~\ref{fig2}(a). The low-energy physics of this plane is well captured by the Emery three-band model \cite{Emery_PhysRevLett.58.2794}, which involves the Cu $3d_{x^2-y^2}$ orbital and the O $p_x$ and $p_y$ orbitals located along the in-plane Cu–O bonds.
Following the standard convention \cite{Zhang_Rice,Emery_PhysRevLett.58.2794}, we define the vacuum $|vac\rangle$ as the fully filled Cu $d^{10}$ and O $p^6$ states from the hole representation. Then, the three-band Hamiltonian \cite{Emery_PhysRevLett.58.2794,Zhang_Rice} can be written as
\begin{eqnarray}
H&=& \Delta_{CT} \sum_{l} p_{l\sigma}^\dagger p_{l\sigma} +U_d \sum_{i} \hat{n}_{i\uparrow}^d\hat{n}_{i\downarrow}^d+U_p \sum_{l} \hat{n}_{l\uparrow}^p\hat{n}_{l\downarrow}^p\nonumber \\
 &+&\sum_{\langle il \rangle} t_{pd}(d_{i\sigma}^\dagger p_{l\sigma}+h.c.)
 -\sum_{\langle ll' \rangle} t_{pp}(p_{l\sigma}^\dagger p_{l'\sigma}+h.c.) 
\end{eqnarray}
where $d_{i\sigma}^{\dagger}$, $p_{l\sigma}^{\dagger}$ are the hole creation operator with spin $\sigma$ for Cu 3$d_{x^2-y^2}$ at site $i$ and O $(2p_{x},2p_{y})$ at site $l$. $t_{pd}$ and $t_{pp}$ are hopping integrals involving $p-d$ and $p-p$ processes respectively, where the angular-dependent hopping signs are defined in Fig.~\ref{fig2}(a). 
The charge-transfer gap is defined as $\Delta_{CT} = \varepsilon_p - \varepsilon_d$, representing the energy difference between the onsite energies of the oxygen $p$ orbital ($\varepsilon_p$) and the copper $d$ orbital ($\varepsilon_d$) in the hole representation.
$U_d$ and $U_p$ are the Hubbard repulsion energies for $d$ and $p$ orbitals respectively. 

Numerous theoretical studies have employed various approaches to estimate the parameters of the three-band Hubbard model relevant to cuprates \cite{Maekawa_JPSJ.59.1760,Emery_PhysRevB.38.4547,Mila_PhysRevB.38.11358,sawatzky_PhysRevB.41.288,andersen1995lda,Martin_PhysRevB.42.6268,Hybertsen_PhysRevB.41.11068,Maekawa_PhysRevLett.66.1228,kent_PhysRevB.78.035132,Arrigoni_2009,Kotliar_PhysRevB.82.125107,vmc_imada3band,PhysRevB.106.235150,PhysRevX.15.021071}. In the literature, typical parameter choices include $U_d \sim 8$–$9$ eV, charge-transfer gap $\Delta_{CT} \sim 3$ eV, and $t_{pp} \sim 0.5 t_{pd}$. However, the value of $t_{pd}$ varies significantly across different works. 
To resolve this ambiguity and determine a more accurate parameter set, we employ a direct fitting approach based on DQMC simulations combined with stochastic analytical continuation. We extract the spin-wave dispersion by locating the peak of the DQMC result of dynamic structure factor, \(S(\mathbf{q},\omega)\), at each momentum \(\mathbf{q}\). 
As shown in Fig.~\ref{fig2}(b), the resulting dispersion reproduces the inelastic neutron scattering data of La$_2$CuO$_4$ with excellent accuracy \cite{La2CuO4_PhysRevLett.86.5377,La2CuO4_PhysRevB.45.7430,Nd2CuO4_PhysRevLett.79.4906}.
To further quantify this agreement, we fit our dispersion to linear spin-wave theory to extract an effective exchange coupling. 
This yields $J \approx 130$~meV, a value nearly consistent with that derived from experimental measurements \cite{La2CuO4_PhysRevLett.86.5377}. 
This agreement in both the overall dispersion and the effective coupling confirms that our chosen parameter set provides a faithful low-energy description of the magnetic properties of cuprates. 

The parameters from this spin-wave fitting is: 
\begin{eqnarray}
     &&t_{pd} = 1.1~\text{eV}, \quad t_{pp} = 0.55~\text{eV}, \quad  \Delta_{CT} = 3.3~\text{eV} \nonumber \\
     && \qquad \qquad U_d = 8.8~\text{eV}, \quad  U_p=0~\text{eV}. 
     \label{parameter}
\end{eqnarray}
Here, we set $U_p = 0$, as this parameter has a negligible impact on the properties under consideration as we will illustrate later. Furthermore,  we should note that a direct comparison of these model parameters with those from density functional theory (DFT) is complicated by the application of particle-hole transformations, which can affect the interpretation of the parameters.
Notably, similar parameters were also employed by Steven R. White and D. J. Scalapino \cite{White_PhysRevB.92.205112}. This validated parameter set will be used for all subsequent calculations unless otherwise specified.



\section{Computational Methods}
\label{sec:methods}

In this work, we employ a multi-pronged computational strategy to comprehensively investigate the magnetic properties of the three-band model. Our approach combines several state-of-the-art numerical techniques, allowing for cross-checking of our results and providing physical insights from different perspectives. This section provides a concise overview of each method, including VMC, DQMC, CP-AFQMC, DMET, and GA, and focuses on how they are utilized to extract key magnetic observables. Further technical details are available in the Appendices.

The primary observables of interest are the local magnetic moments and the spin structure factor. The local moment at site $i$ with orbital $\mu \in \{d,p_x,p_y\}$ is defined as 
\begin{equation}
m_{i\mu} \equiv \frac{1}{2}\left\langle \hat{n}^\mu_{i\uparrow} - \hat{n}^\mu_{i\downarrow} \right\rangle.    \label{eq:local_moment}
\end{equation}
The static spin structure factor at momentum ${\bf q}$ is given by 
\begin{equation}
\begin{split}
&S({\bf q})=\sum_{a \in \{x,y,z\}} \langle S^a({\bf q})S^a(-{\bf q})\rangle,\quad 
S^{a}({\bf q})=\sum_{i} \frac{e^{-i{\bf q}\cdot{\bf r}_{i}}}{N_c} S_{i}^{a},
\end{split}
\label{eq:SQ_def}
\end{equation}
where $S_{i}^{a}=\frac{1}{2}\sum_{\mu\sigma\sigma'}c_{i\mu\sigma}^{\dagger}\sigma_{\sigma\sigma'}^{a}c_{i\mu\sigma'}$ is the total spin operator in unit cell $i$, $N_c=L^2$ is the number of unit cells, and ${\bf r}_i$ are their positions. The AFM order parameter $m_z$ is then estimated from the structure factor at the ordering vector ${\bf Q}=(\pi, \pi)$ via $m_z=\sqrt{S({\bf Q})}$~\cite{vmc_imada3band,vmc_yamaji2022}.

To further characterize the magnetic properties beyond the static features provided by the spin structure factor, static spin susceptibility, which incorporates thermal fluctuations, is computed using DQMC. It is defined for each spin component ($a=x,y,z$) as:
\begin{equation}
\chi^{a}(\mathbf{q}) =
\int_{0}^{\beta} d\tau
\langle S^{a}({\bf q},\tau) S^{a}({-\bf q},0) \rangle,
\label{eq:chi}
\end{equation}
where $S^{a}({\bf q}, \tau) = e^{\tau H} S^{a}({\bf q}) e^{-\tau H}$ is the spin operator in the imaginary-time Heisenberg picture, with $S^{a}({\bf q})$ defined in Eq.~\eqref{eq:SQ_def}. This quantity is a crucial probe of the system's spin fluctuations and tendency towards magnetic order, as its divergence can signal a continuous phase transition.

\subsection{Variational Monte Carlo}

Variational Monte Carlo (VMC) is a powerful method for determining ground-state properties by optimizing a parameterized trial wave function. The trial wave function, or ansatz, is typically constructed as $\left|\Psi\right\rangle = \mathcal{P}\left|\Phi_{0}\right\rangle$. Here, $\left|\Phi_{0}\right\rangle$ is a mean-field state (e.g., a Slater determinant or Pfaffian) that captures the system's band structure, and $\mathcal{P}$ is a set of projectors incorporating strong correlation effects. The variational parameters, contained within both the mean-field state $|\Phi_{0}\rangle$ and the projectors, are optimized to minimize the total energy, providing a systematically improvable upper bound to the true ground-state energy. This optimization is performed efficiently using stochastic reconfiguration updates~\cite{vmc_sorella_1,vmc_sorella_2}.

To balance accuracy with computational feasibility, we employ two distinct VMC schemes in this work. The first is the many-variable VMC (mVMC) method~\cite{vmc_MISAWA2019,vmc_Imada2016}, which utilizes a highly elaborated ansatz: $\left|\Psi\right\rangle =\mathcal{P}_{G}\mathcal{P}_{J}\mathcal{P}_{S}\left|\phi_{\mathrm{pair}}\right\rangle$. This wave function combines Gutzwiller projector $\mathcal{P}_{G}$~\cite{vmc_Gutzwiller}, Jastrow factor $\mathcal{P}_J$~\cite{vmc_Jastrow}, and spin-projection $\mathcal{P}_S$ operators~\cite{vmc_Imada2008,vmc_quantumProj} with a Bogoliubov-pairing Pfaffian state $\left|\phi_{\mathrm{pair}}\right\rangle$. While offering a more accurate description of the ground state, its large parameter space make it computationally demanding. 

Crucially, we find that the results from the mVMC method are in quantitative agreement with those from a less demanding scheme, e.g., the conventional VMC. For this reason, we adopt the latter for calculations on larger systems and for studying spatial inhomogeneity. This conventional VMC scheme employs a simpler ansatz, $\mathcal{P}_{G}\left|\phi_{\mathrm{mf}}\right\rangle$, where the mean-field state $|\phi_{\mathrm{mf}}\rangle$ simply incorporates an AFM order parameter. This simpler form also offers more direct physical intuition. Using both schemes, we evaluate ground-state properties, primarily focusing on the spin structure factor to characterize the magnetic order.

\subsection{Determinant Quantum Monte Carlo}

In contrast to VMC, DQMC is an unbiased algorithm~\cite{DQMC_PhysRevD.24.2278,DQMC_PhysRevLett.46.519,DQMC_PhysRevLett.47.1628,AssaadReview,Li2019review}, which stochastically evaluates the thermal expectation value of an observable \(\hat{O}\) within the grand canonical ensemble:
$\langle \hat{O} \rangle =
\frac{{\rm Tr}\left[\hat{O} e^{-\beta (\hat{H} - \mu \hat{N})}\right]}
{{\rm Tr}\left[e^{-\beta (\hat{H} - \mu \hat{N})}\right]}.$
Here \(\beta\) is the inverse temperature and \(\mu\) is the chemical potential for the grand canonical ensemble. 
The evaluation of the trace proceeds by first discretizing the imaginary-time axis $\beta$ into $N_{\tau}$ slices of duration $\Delta\tau = \beta/N_{\tau}$, using a symmetric second-order Trotter-Suzuki decomposition. This introduces a systematic error that vanishes as $\Delta\tau \to 0$. In practice, we choose a sufficiently small time step, $\Delta\tau = 0.1$, to ensure this error is negligible. Subsequently, the quartic Hubbard interaction term is decoupled via a discrete Hubbard-Stratonovich (H-S) transformation, mapping the Hubbard model onto a system of non-interacting fermions coupled to a fluctuating auxiliary field $\{s\}$. The fermion part is traced out analytically, arriving at the weight $W_{\{s\}}$ depending on the configuration of auxiliary fields. 

A significant challenge in the approach of DQMC is that the Monte Carlo weight, $W_{\{s\}}$, is not guaranteed to be positive definite, giving rise to the infamous fermion sign problem~\cite{White1990PRBsignproblem,Wu2005PRB,Yao2016PRL}. Here, to circumvent this, the stochastic sampling is performed using the absolute value of the weight, $|W_{\{s\}}|$. 
The physical expectation value of an observable $\hat{O}$ is then recovered by reintroducing the sign of the weight into the measurement: $
\langle \hat{O} \rangle = \frac{\langle \hat{O} \cdot \mathrm{sgn}(W_s) \rangle_{|W_s|}}{\langle \mathrm{sgn}(W_s) \rangle_{|W_s|}}$.
The denominator of this expression, $\langle \mathrm{sgn}(W_s) \rangle_{|W_s|}$, is the average sign. 
For generic models of interest, the average sign decays exponentially with increasing system size and inverse temperature $\beta$~\cite{Troyer2005PRLsign}. This exponential decay requires an exponentially large number of Monte Carlo samples to resolve the numerator against the statistical noise, severely limiting the accessible parameter regimes of temperature and system size. While this restricts our simulations to finite systems and relatively high temperatures, it is sufficient to characterize the essential magnetic properties and the system's ordering tendencies.

The dynamic spin structure factor \(S^{z}(\mathbf{q},\omega)\) is calculated using DQMC, as it is directly related to the inelastic neutron scattering experiments. Rather than real-frequency spectrum,  the DQMC simulations, which is carried out in imaginary time, yields imaginary-time spin correlation function \(G(\mathbf{q},\tau) = \langle S^{z}({\mathbf{q}},\tau) S^{z}({-\mathbf{q}},0)\rangle\) only. Nevertheless, the connection between these two domains is established through the spectral representation:
\begin{equation}
G(\mathbf{q},\tau)
= \int_{0}^{\infty} \frac{d\omega}{\pi}
S^{z}(\mathbf{q},\omega)
\left( e^{-\tau\omega} + e^{-(\beta-\tau)\omega} \right),
\label{eq:Gtau}
\end{equation}
 The task of extracting \(S^{z}(\mathbf{q},\omega)\) requires inverting this integral transform, which is a notoriously ill-posed problem. To overcome this challenge, we employ the well-established stochastic analytical continuation method~\cite{Sandvik1998PRBSAC}.

\subsection{Constrained-Path Auxiliary-Field Quantum Monte Carlo}

CP-AFQMC~\cite{PhysRevB.55.7464,PhysRevLett.83.2777} was developed for interacting fermion systems which suffer from the infamous sign problems~\cite{PhysRevB.41.9301,PhysRevLett.94.170201}.
Similar to the DQMC, in CP-AFQMC, ground state is obtained by imaginary-time evolution in the large time limit. The kinetic and interaction part in the imaginary-time projector $e^{-\tau H}$ is decoupled via a Trotter-Suzuki decomposition, and the interaction term is decoupled using a Hubbard-Stratonovich transformation. Instead of tracing out the fermion part in DQMC, CP-AFQMC deal directly with the wave-function. The projected wave function $|\psi\rangle$, which is evolved by $e^{-\tau H}$, can be expressed as a weighted sum over a population of walkers $\{|\phi_n\rangle\}$ as $|\psi\rangle = \Sigma_n w_i|\phi_n\rangle\ $, where each walker is a single Slater determinant. To control the sign problem, a trial wave function $|\psi_T\rangle$ is employed. 
The trial wave function serves as a guide in the  imaginary-time evolution by removing the walkers with negative weight $w_n$ from the simulations. This procedure get rid of the minus sign problem with the price of a systematic bias. The magnitude of this bias is determined by the quality of the trial wave function.





In this work, to systematically minimize the bias inherent in CP-AFQMC, we optimize the trial wave-function self-consistently \cite{PhysRevB.94.235119,PhysRevB.107.235124}. This iterative procedure begins with an initial trial wave function and the trial wave-function is optimized until convergence of physical quantity is achieved. A brief introduction to the two self-consistent approaches using mean-field Hamiltonian \cite{PhysRevB.94.235119} and  one-body reduced density matrices from mixed estimators \cite{PhysRevB.107.235124} can be found in the Appendix. 

We measure the AFM order in CP-AFQMC by introducing an AFM pinning field~\cite{PhysRevX.3.031010}. This field is applied along a single edge of the lattice, defined by unit cells ${\bf r}_i=(1, y_i)$, which gives an addition term to the Hamiltonian:
\begin{equation}
H_h = h \sum_{{\bf r}_i=(1, y_i)} e^{i \mathbf{Q} \cdot \mathbf{r}_i} S_{i}^z
\end{equation}
where $h$ is the field strength (set as $h=0.2$) and $\mathbf{Q}=(\pi,\pi)$.
We measure the staggered magnetization on the edge which has the largest distance to the pinning edge. For a system with PBC, this corresponds to the edge of sites at ${\bf r}_i=(1+L/2, y_i)$. For a given size, The AFM order parameter is calculated as:
\begin{equation}
m_z =\sum_{{\bf r}_i=(1+L/2, y_i)} e^{i \mathbf{Q}\cdot\mathbf{r}_i} \left\langle S_{i}^z\right\rangle / L
\end{equation}
The spontaneous magnetization in the thermodynamic limit is then determined by extrapolating these results to $L \to \infty$.

In the hole doped case, the converged results from the self-consistent optimization doesn't give AFM state but stripe state \cite{PhysRevB.102.214512}. To estimate the magnetization in the hole doped side by forcing the ground state in the AFM manifold, we intentionally used AFM-type trial wave-function and optimize the magnitude of the AFM order in the trial wave-function by minimizing the QMC energy. More details can be found in the Appendix. We also perform a comparison of this approach and the self-consistent approach in the electron doped case where the more accurate self-consistent calculations indeed gives the AFM ground state and find a good agreement between them (see the results in the next section). 

\subsection{Density Matrix Embedding Theory}

DMET is a quantum embedding framework that maps the ground state of a large system onto a smaller, computationally tractable impurity problem through a Schmidt decomposition~\cite{DMET_PhysRevLett.109.186404,Schmidt_decomp}. Based on the recent application of DMET to the three-band model~\cite{DMET_PhysRevResearch.2.043259}, we embed a \(C_4\) symmetric \(2\times 2\) \(\mathrm{CuO}_2\) (\(N_\text{orb}=12\)) cluster as the impurity into a larger $40\times 40$ lattice. DMET maps a large \(L\times L\) lattice (\(L=40\)) onto an effective impurity problem consisting of \(N_\text{orb}\) correlated (impurity) sites and an equal number of bath orbitals. 

Since the exact ground state is unknown, the embedding is constructed from an auxiliary mean-field Hamiltonian~\cite{DMET_guide}:
\begin{equation}
\hat{h}
= \hat{h}_{0} + \hat{u}
= \hat{h}_{0} + \sum_{i j \sigma} v_{i j}^{\sigma}\, c_{i\sigma}^{\dagger} c_{j\sigma},
\label{eq:correlation_potential}
\end{equation}
where \(\hat{h}_{0}\) is the non-interacting part of the full Hamiltonian \(\hat{H}\) and $c_{i\sigma}$ represents an electron in either a Cu-$d$ or O-$p$ orbital.
The correlation potential $\hat{u}$ [e.g., the parameter $v_{i j}^{\sigma}$ in Eq.~\eqref{eq:correlation_potential}] is optimized to make the embedded system match the full system. To capture the magnetic properties of the three-band model, we allow \(\hat{u}\) to break spin SU(2) down to U(1) along the \(z\) axis. From the mean-field solution, we construct the embedding problem, which we solve using the coupled-cluster singles and doubles (CCSD) method~\cite{CCSD_method}. With the help of DMET, the local magnetic moment in Eq.~\eqref{eq:local_moment} is calculated efficiently.

\subsection{Gutzwiller Approximation}

The Gutzwiller projection offers an efficient way to account for the suppression of double occupancy in correlated electron systems~\cite{GW_PhysRev.137.A1726,GW_PhysRevLett.10.159}. While exact projection is demanding, its effect can be captured qualitatively  at a mean-field level via the Gutzwiller Approximation (GA). Our calculations utilize the Kotliar–Ruckenstein slave-boson formalism, which is known to be equivalent to the GA in the saddle-point approximation~\cite{KR_PhysRevLett.57.1362}.

In the slave-boson formalism, the physical electron operator for each orbital $\alpha$ is fractionalized into a pseudo-fermion ($\hat{f}_{i\sigma}^\alpha$) and auxiliary bosons that track the local charge configurations. The local Hilbert space is expanded to include a holon ($\hat{e}_i^\alpha$), a doublon ($\hat{d}_i^\alpha$), and spinon bosons ($\hat{p}_{i\sigma}^\alpha$). 
The local physical states are constructed in an enlarged space as 
\begin{equation}
\begin{split}
\vert \text{empty} \rangle_i^\alpha &= \hat{e}_i^{\alpha\dagger} \vert \text{vac} \rangle, \\
\vert \sigma \rangle_i^\alpha &= \hat{p}_{i\sigma}^{\alpha\dagger} \hat{f}_{i\sigma}^{\alpha\dagger} \vert \text{vac} \rangle, \\
\vert \uparrow\downarrow \rangle_i^\alpha &= \hat{d}_i^{\alpha\dagger} \hat{f}_{i\uparrow}^{\alpha\dagger} \hat{f}_{i\downarrow}^{\alpha\dagger} \vert \text{vac} \rangle.
\end{split}    
\end{equation}
To ensure this representation remains faithful to the original physical Hilbert space, two local constraints are imposed, such as
\begin{equation}
\begin{split}
&\hat{e}_i^{\alpha\dagger} \hat{e}_i^{\alpha} + \sum_{\sigma} \hat{p}_{i\sigma}^{\alpha\dagger} \hat{p}_{i\sigma}^{\alpha} + \hat{d}_i^{\alpha\dagger} \hat{d}_i^{\alpha} = 1 \\
&\hat{f}_{i\sigma}^{\alpha\dagger} \hat{f}_{i\sigma}^{\alpha} = \hat{p}_{i\sigma}^{\alpha\dagger} \hat{p}_{i\sigma}^{\alpha} + \hat{d}_i^{\alpha\dagger} \hat{d}_i^{\alpha}
\end{split}    
\end{equation}
The benefit of this formalism is apprant by noting that the on-site Coulomb interaction simplifies to $U_{\alpha} \hat{n}_{i\uparrow}^\alpha \hat{n}_{i\downarrow}^\alpha \to U_{\alpha} \hat{d}_i^{\alpha \dagger} \hat{d}_i^{\alpha}$. The kinetic hopping term, however, is renormalized by the GA factor $z_{i\sigma}^{\alpha\dagger} z_{j\sigma}^{\beta}$, where the $z$ operators are complex functions of the bosons that correctly manage the transitions between different charge configurations~\cite{KR_PhysRevLett.57.1362}.

In the mean-field approximation, the boson operators are replaced by their expectation values. The ground state is then found by self-consistently minimizing the energy with respect to these bosonic mean-field parameters. The local magnetic moment is subsequently calculated from the pseudo-fermions as
\begin{equation}
m_i^\alpha = \frac{1}{2} \langle \hat{f}_{i\uparrow}^{\alpha\dagger} \hat{f}_{i\uparrow}^{\alpha} - \hat{f}_{i\downarrow}^{\alpha\dagger} \hat{f}_{i\downarrow}^{\alpha} \rangle.
\end{equation}


\subsection{Benchmark}
\label{sec:benchmark}

\begin{table}
\begin{centering}
\begin{tabular}{|c|c|c|c|c|c|c|}
\hline 
Methods &  GA & DMET & VMC & mVMC & CP-AFQMC \tabularnewline 
\hline 
$m_{z}$ & 0.297 & 0.286 & 0.295 & 0.286 & 0.2404\tabularnewline
\hline 
$\pm\delta m_z$ & \multicolumn{1}{c|}{\diagbox[width=0.05\linewidth]{}{}} & $1\times10^{-4}$ & 0.0016 & 0.0065 & $5.4\times10^{-4}$ \tabularnewline
\hline
\end{tabular}\caption{Thermodynamic-limit AFM order parameter $m_{z}$ at $x=0$ for the three-band Emery model, estimated by various numerical methods.}\label{table_AFM}
\par
\end{centering}
\end{table}

\begin{figure}[t]
  \centering
  \includegraphics[width=0.47\textwidth]{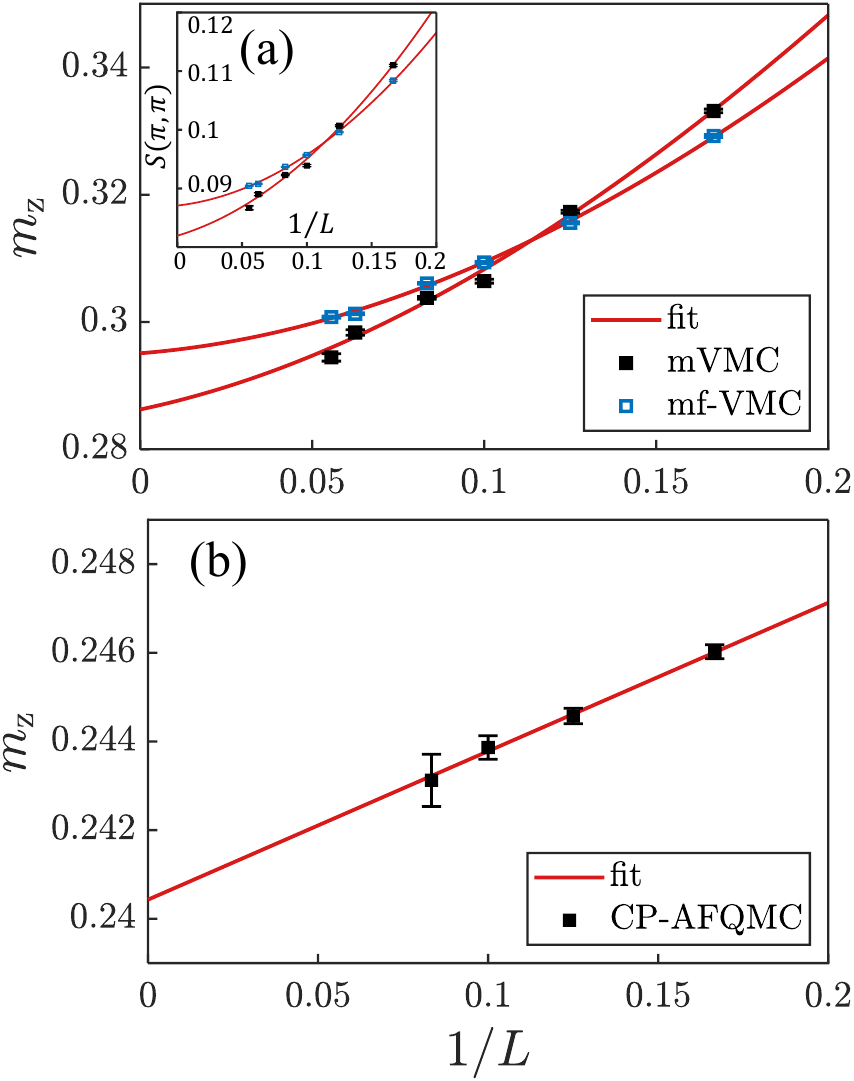}
  \caption{Undoped $(x=0)$ Anti-ferromagnetic magnetization in the three-band Hubbard model. (a) VMC-based results: peak spin structure factor $S(\pi,\pi)$ versus $1/L$; a quadratic fit in $1/L$ is used to extrapolate to the thermodynamic limit (TL), from which the staggered moment  $m=\sqrt{S(\pi,\pi)}$ is obtained. (b) CP-AFQMC: induced staggered magnetization measured at the sites farthest from the pinning centers versus $1/L$. A linear fit with  $1/L$ yields the TL estimate.}
  \label{fig3}
\end{figure}

This subsection provides a benchmark study to assess the consistency and limitations of different computational approaches. All simulations are performed on finite-sized clusters with periodic boundary conditions. 
Unless otherwise specified, we adopt the hole representation customary for cuprates, wherein the undoped, half-filled state contains one hole per CuO$_2$ unit cell ($n_h=1$). A system with hole doping $x$ therefore has a total hole density of $n_h = 1+x$ and positive (negative) $x$ indicates hole (electron) doping.

We begin by examining the most accessible quantity---the AFM order parameter at $x = 0$ in the thermodynamic limit. Among the methods considered, GA and DMET are both mean-field–type approaches in which $m_z$ can be obtained directly in the infinite-system limit. Their self-consistent solutions yield $m_z = 0.297$ and $m_z = 0.286$, respectively.

For Monte Carlo–based approaches, the thermodynamic limit must be reached by finite-size extrapolation.
Using the VMC-type methods (VMC and mVMC), we evaluate the AFM order through the spin structure factor peak $S(\mathbf{Q})$ and fit it as a quadratic function of $1/L$, as illustrated in the inset of Fig.~\ref{fig3}(a) \cite{vmc_MISAWA2019}.
The corresponding $m_z$ values, extracted from $S(\mathbf{Q})$, are shown in Fig.~\ref{fig3}(a), yielding $m_z \approx 0.295$ for VMC and $m_z \approx 0.286$ for mVMC---both in close agreement with the GA and DMET results.

As mentioned above, to determine the spontaneous AFM order within CP-AFQMC, the boundary-pinning method~\cite{PhysRevX.3.031010} is employed. We perform a finite size scaling to determine the spontaneous AFM order in the thermodynamic limit. As shown in Fig.~\ref{fig3}(b), this approach yields $m_z \approx 0.24$,
which is suppressed relative to our VMC and mean-field estimates, reflecting the stronger quantum fluctuations captured by the CP-AFQMC method.

To determine the ground-state staggered magnetization $m_z$ with DQMC, we compute the spin structure factor $S(\mathbf{Q})$ for various system sizes $L$ at finite temperature. To reliably access the ground state, we use an inverse temperature $\beta=2L$ that scales with the system size. The thermodynamic value of $m_z$ is then obtained by performing a finite-size scaling extrapolation of $\sqrt{S(\mathbf{Q})}$, a procedure detailed in the Appendix. This analysis yields $m_z \approx 0.245$, a result in excellent agreement with our CP-AFQMC calculations. A summary of the AFM order parameters obtained from all methods in the thermodynamic limit is provided in Table~\ref{table_AFM}.
The obtained $m_z$ is consistent with experimental observation around $0.2 \sim 0.3$ in cuprates~\cite{schrieffer2007handbook}.

\begin{figure*}[t]
  \centering
  \includegraphics[width=0.98\textwidth]{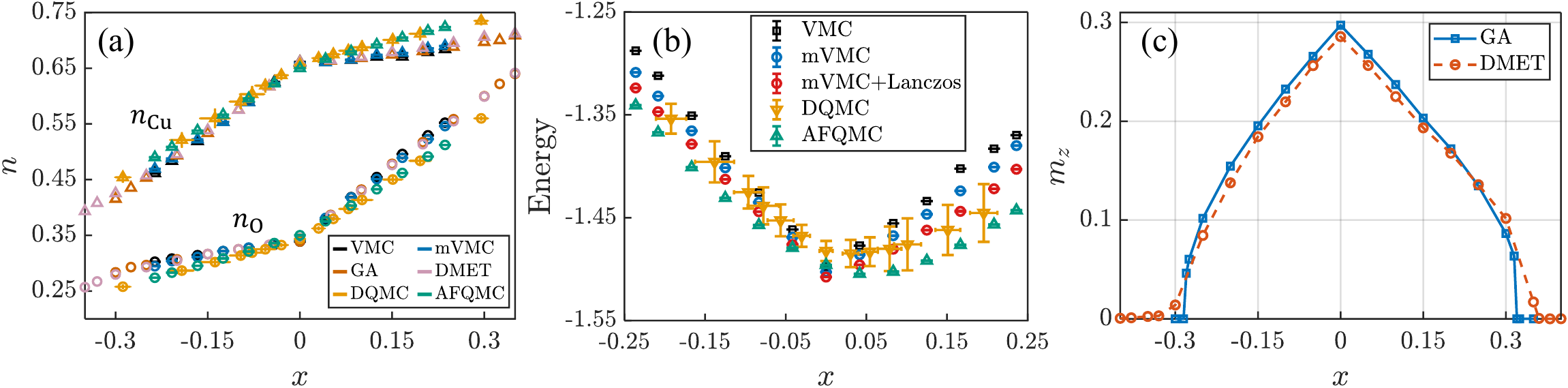}
  \caption{Orbital-resolved densities, energies and AFM moments from various methods on $L=12$ clusters. (a) Triangles denote the Cu-site density $n_{\mathrm{Cu}}$; circles denote the oxygen-site density $n_{\mathrm{O}}$ (sum over $p_x$, $p_y$). (b) Ground-state energy per unit cell versus doping $x$. (c) AFM moment from GA and DMET ($2\times2$ impurity cluster with a CCSD solver). All data use the same model parameters; see Appendix for method-specific details.}
  \label{fig_benchmark}
\end{figure*}

\section{Magnetic Electron-hole phase diagram}
\label{sec:phase-diagram}

Before analyzing the AFM order away from half-filling, we perform a necessary benchmark by comparing key observables across our computational methods, focusing on a $12 \times 12$ lattice.  
Fig.~\ref{fig_benchmark}(a) shows the orbital-resolved hole densities versus doping. 
Across the entire accessible doping range, quantitative agreement is found among all methods for the charge distributions. Moreover, our results reveal a pronounced asymmetry between the hole-doped ($x>0$) and electron-doped ($x<0$) regimes.
Upon hole doping, the oxygen hole density $n_\text{O}$ increases much more rapidly than the copper hole density $n_\text{Cu}$, whereas this trend is reversed under electron doping. This asymmetry reflects the characteristic behavior of charge-transfer systems, where doped holes predominantly occupy oxygen orbitals, while doped electrons primarily reside on copper sites.

The energy obtained from different methods on the $L=12$ lattice are shown in Fig.~\ref{fig_benchmark}(b). For clarity, we exclude the GA and DMET results and focus on the various Monte Carlo approaches. The DQMC data are computed at an inverse temperature of $\beta = 8$, while the VMC, mVMC, and CP-AFQMC results are zero-temperature estimates. 
We can find that all the methods agree well on the electron-doping side. On the other hand, at larger $x$, the VMC energies deviate slightly upward. This deviation originates from the restricted variational ansatz (limited number of parameters) and the omission of competing fluctuations beyond the AFM order, both of which lead to a systematic overestimation of the ground-state energy $E_0(x)$.

Combining these computational approaches, we now address the central goal of this work---exploring the magnetic phase diagram for both electron- and hole-doped regimes. We begin with the GA and DMET methods, which determine the AFM order parameter from their respective self-consistent mean-field solutions. 
As shown in Fig.~\ref{fig_benchmark}(c), both methods produce highly consistent results, namely, the comparable magnetic moments. Moreover, the boundaries of the AFM ordered ground state are similar with respect to the doping level $x$: $x \in [-0.285,0.32]$ for GA and $x \in [-0.32,0.36]$ for DMET. This indicates that no pronounced electron–hole asymmetry is observed---the maximum hole-doping range is even slightly larger than that of the electron side. This trend is consistent with the previous DMET results \cite{DMET_PhysRevResearch.2.043259}.
Although mean-field approaches are known to overestimate magnetic order, we expect the overall trend to persist once quantum fluctuations are included. And these mean-field phase boundaries can be regarded as upper bounds for the stability of AFM.

\begin{figure*}[t]
  \centering
  \includegraphics[width=0.98\textwidth]{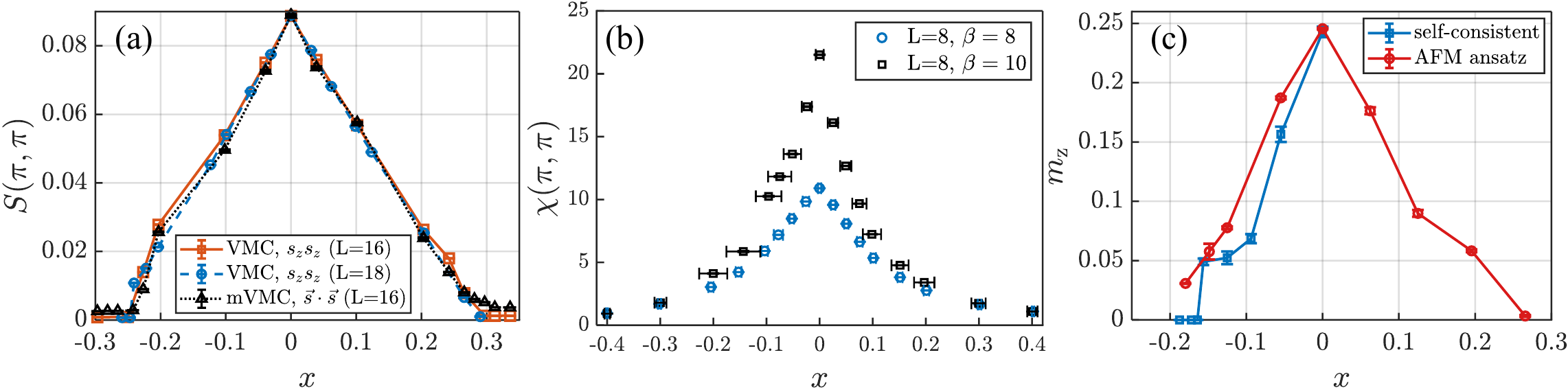}
  \caption{(a) The spin correlation function at momentum $Q=(\pi,\pi)$ as a function of doping $x$ estimated by VMC ($L=16,18$) and mVMC ($L=16$, no spin quantum-number projection). (b) DQMC uses $L=8$, $\beta=8,10$ to obtain the staggered spin susceptibility. (c) CP-AFQMC calculated stagger magnetization $m_z$ on $L=16$, using two approaches. For the blue line trial wave-functions are optimized self-consistently while for the red line an optimized AFM trial wave-function is used to force AFM order. When the self-consistent calculation indeed gives AFM state, the two methods agree qualitatively well. For the hole doped side where self-consistent calculation gives stripe state, we only show the results of second approach. All methods show no discernible electron–hole asymmetry; a mild hole-side enhancement is suggested. See Appendix for computational details.}
  \label{fig4}
\end{figure*}

To examine the phase boundary beyond mean-field level, we employ the VMC and mVMC methods. At each doping level $x$, we identify the ground state by independently optimizing the trial wave functions initialized in the AFM and paramagnetic sectors and selecting the one with lower energy. The AFM boundary is determined from the energy crossing between these two solutions, while static spin structure factors are extracted accordingly. 
As shown in Fig.~\ref{fig4}(a), the staggered spin correlations decrease approximately linearly with ($|x|$) on both the electron- and hole-doped sides, with slightly better linearity observed for hole doping. Using the AFM–PM energy crossing as the criterion, we estimate the apparent AFM region to be $x\approx[-0.242,\,0.29]$. Thus, consistent with previous mean-field results, the VMC calculations also show no clear electron–hole asymmetry.

While the results in Fig.~\ref{fig4}(a) are not extrapolated to the thermodynamic limit, we have confirmed that the phase boundary are robust against finite-size and/or finite-temperature effects. For instance, the static spin structure factors between $L=16$ and $L=18$ systems exhibit excellent agreement, 
indicating negligible finite-size effects.
Moreover, the results from mVMC and VMC at ($L=16$) differ only slightly, suggesting that the simpler VMC ansatz already provides a reliable estimate of the AFM boundary.

Furthermore, to directly compare the AFM ordering tendency, we employ DQMC to evaluate the finite-temperature staggered spin susceptibility, $\chi(\pi,\pi)$. An enhancement of $\chi(\pi,\pi)$ with decreasing temperature signals the AFM correlation, while its suppression with doping marks the destruction of the order. While finite-temperature DQMC alone cannot precisely determine the AFM phase boundary, it enables a direct comparison between electron and hole doping, as shown in Fig.~\ref{fig4}(b).
For instance, at $x = 0.2$ and $x = -0.2$, the values of $\chi(\pi,\pi)$ remain comparable for both $\beta=8$ and $\beta=10$.
Thus, the spin susceptibility exhibits no clear electron–hole asymmetry, reinforcing the conclusion drawn from our other methods.


CP-AFQMC provides a direct AFM order parameter $m_z(x)$ through the pinning field approach in the electron doped side, whose disappearance signals the loss of long-range order. 
In Fig.~\ref{fig4}(c), CP-AFQMC results with the AFM mean-field trial wave function to force AFM ground state gives similar results as those from the more accurate self-consistent approach in the electron doping side. 

In the hole doped side where self-consistent approach can't stabilize the AFM order and give stripe instability \cite{PhysRevB.102.214512}, we show the results from the calculation with optimized AFM-type trial wave-function to force a AFM state in CP-AFQMC. In Fig.~\ref{fig4}(c), we can find the AFM magnetizations are comparable between the electron and hole doped sides. Taken together, the qualitative conclusions above remain robust, though thermodynamic extrapolation is not performed.

Overall, despite utilizing different observables and systematics, all methods paint a consistent picture: no resolvable electron–hole asymmetry within uncertainties, and even a slightly stronger AFM tendency on the hole side if we only consider AFM order and ignore other competing orders such as stripe \cite{YANAGISAWA2002292,White_PhysRevB.92.205112,PhysRevB.102.214512,PhysRevB.108.205154,Huang2017Science,doi:10.1073/pnas.2408717121}.

\subsection{$U_p$ dependence}

To assess the sensitivity of magnetism to correlations on the oxygen sites, we compare the baseline $U_p=0$ calculations with data at $U_p=2$. We analyze two complementary observables. First, VMC on $L=16$ clusters tracks the staggered spin correlations as a function of doping $x$. Second, we use DQMC to calculate the finite-temperature staggered susceptibility, $\chi(\pi,\pi)$. However, a non-zero $U_p$ exacerbates the fermion sign problem, restricting these DQMC simulations to smaller $L=8$ clusters at an inverse temperature of $\beta=8$.

\begin{figure}[t]
  \centering
  \includegraphics[width=0.46\textwidth]{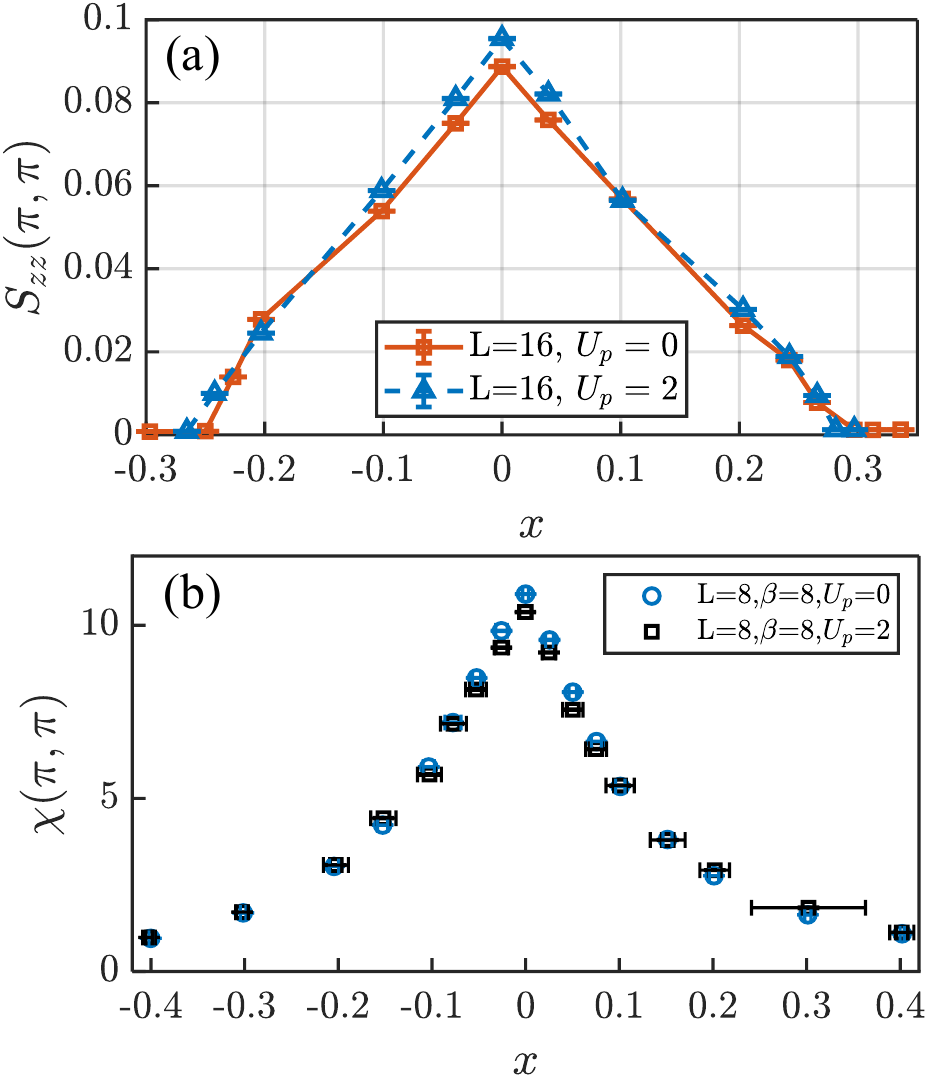}
  \caption{(a) The spin correlation function for various $U_p$ parameters as a function of doping $x$ estimated by VMC. (b) Spin susceptibility calculated by DQMC.}
  \label{fig_Up}
\end{figure}

As shown in Fig.~\ref{fig_Up}, both probes point to the same conclusion: within our accessible sizes and temperatures, introducing $U_p=2$ yields only minor changes in the AFM tendency across dopings. In VMC, the $x$-dependence of the staggered spin correlations nearly coincides with the $U_p=0$ curves, with a slight enhancement near $x=0$. In DQMC, $\chi(\pi,\pi)$ for $U_p=2$ largely overlaps the $U_p=0$ result, but shows a small suppression near $x=0$, which is the opposite trend to VMC. Disentangling potential $U_p$ effects from finite-$T$ bias requires a controlled $\beta$-dependence at fixed $L$, which is beyond the present scope. Consequently, the qualitative conclusion established above—most notably the absence of an electron–hole asymmetry and a mild indication of stronger magnetism on the hole side—remains unchanged upon including $U_p=2$.

We emphasize that these comparisons are performed at fixed finite $L$ (and finite $\beta$ for DQMC) without thermodynamic or zero-temperature extrapolation. Our analysis therefore provides a qualitative comparison and does not aim to determine the precise thermodynamic phase boundary. Even with this caveat, the central finding that the pronounced electron-hole asymmetry is absent remains robust against a moderate oxygen-site repulsion.

\subsection{{Nd$_2$CuO$_4$} VS {La$_2$CuO$_4$}}
In the previous section, we show that the antiferromagnetism in the three-band model exhibits comparable strength on both the electron- and hole-doped sides if we only consider AFM instability. 
One may ask whether material-specific differences can cause asymmetry regarding AFM order? Nd$_2$CuO$_4$ is a prototypical electron-doped cuprate, with $T_N$ and $T_c$ similar to La$_2$CuO$_4$, as discussed in the introduction. In this subsection, we aim to establish a suitable parameter set to describe Nd$_2$CuO$_4$ to check whether this influences our conclusion.

The primary structural difference between Nd$_2$CuO$_4$ and La$_2$CuO$_4$ lies in the presence of apical oxygen atoms. In the infinite-layer structure of Nd$_2$CuO$_4$, the apical oxygens above the Cu sites are absent. This leads to two key effects: (1) an increase in the in-plane lattice constant; (2) a reduction in the charge-transfer gap \cite{Madelung_PhysRevLett.63.1515,Madelung_PhysRevLett.66.1228}. As a result, the hopping parameters $t_{pd}$ and $t_{pp}$, as well as the charge-transfer energy $\Delta_{CT}$, must be adjusted accordingly, while other interaction parameters are kept the same. To estimate the changes in hopping amplitudes, we employ Harrison’s scaling relation \cite{harrison2012electronic}. For example, the Cu–O hopping integral $t_{pd}(r)$ at a bond length $r$ scales with a reference value $t_{pd}^0$ at distance $r_0$ as $(r/r_0)^{-3.5}$ while $t_{pp}$ has a $(-2)$ power scaling relation \cite{harrison2012electronic,sawatzky_PhysRevB.41.288}.

The in-plane lattice constant of La$_2$CuO$_4$ is 3.818 \AA \cite{LaCuO_crystal_PhysRevB.38.11337}, while that of Nd$_2$CuO$_4$ is 3.943 \AA \cite{nco_PhysRevLett.62.1197}, leading to modified hopping parameters of $t_{pd}^e = 1.0$ eV and $t_{pp}^e = 0.5$ eV for Nd$_2$CuO$_4$. 
We then determine the charge-transfer gap $\Delta_{CT}^e$, by reproducing similar experimental spin excitation spectrum. Using the same DQMC and stochastic analytical continuation methodology as for La$_2$CuO$_4$, we find that $\Delta_{CT}^e = 2.8$ eV provides the best fit. This value is also consistent with optical measurements, which report an optical gap difference of approximately $0.4 \sim 0.5$ eV between La$_2$CuO$_4$ and Nd$_2$CuO$_4$ \cite{Optical_PhysRevB.42.10785,Maekawa_PhysRevLett.66.1228,optical_cuo2,optical_PhysRevB.41.11657}. 

\begin{figure*}[t]
  \centering
  \includegraphics[width=0.98\textwidth]{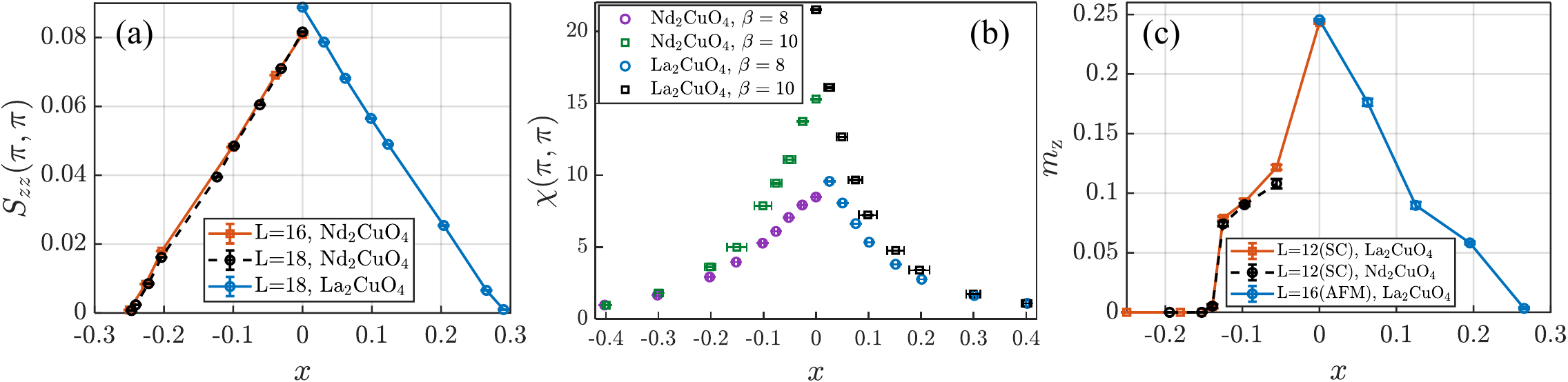}
  \caption{Magnetic response with a parameter set appropriate for electron-doped cuprates. (a) Spin correlations from VMC. (b) Staggered spin susceptibility from DQMC using the same parameters. (c) AFM magnetization from CP-AFQMC with the same parameters; “SC” denotes a self-consistently optimized trial wave function, and “AFM” denotes an AFM trial wave function.}
  \label{fig_Nd}
\end{figure*}

To assess whether material–specific parameters could account for the discrepancy, we recompute the electron-doped AFM phase diagram of the three-band Emery model with parameters suitable for Nd$_2$CuO$_4$. VMC on $L=16,18$ evaluates the staggered spin correlation $S_{zz}(\pi,\pi)$ [Fig.~\ref{fig_Nd}(a)]; DQMC on $L=8$ at $\beta=8,10$ yields the staggered susceptibility $\chi(\pi,\pi)$ [Fig.~\ref{fig_Nd}(b)]; and CP-AFQMC on $L=12$ estimates the AFM order parameter $m_z$ [Fig.~\ref{fig_Nd}(c)]. Across these observables, the magnitude and doping dependence of the AFM response with Nd$_2$CuO$_4$ parameters closely track those obtained with La$_2$CuO$_4$ on the electron-doped side. For a direct comparison to the hole-doped sector, Fig.~\ref{fig_Nd} juxtaposes the Nd$_2$CuO$_4$ (electron side) results with the La$_2$CuO$_4$ (hole side) results: no electron–hole asymmetry is observed within our accessible sizes and temperatures, and the hole side even exhibits a slightly stronger AFM tendency [Fig.~\ref{fig_Nd}(a,c)], consistent with our earlier findings. These comparisons indicate that our conclusions are robust against reasonable variations of the model parameters.

\section{Doping potential}
\label{sec:doping}

As discussed above, our results indicate that the AFM state in the three-band model does not exhibit a clear electron–hole asymmetry if we consider only AFM order. Historically, the greater robustness of AFM in electron-doped cuprates has been attributed to spin-dilution effects \cite{lco_PhysRevB.45.7430}. This interpretation is supported by neutron scattering experiments showing that Zn substitution in La$_2$CuO$4$ suppresses the Néel temperature at a rate comparable to that produced by Ce doping in Pr$_{2-x}$Ce$_x$CuO$_4$ \cite{lco_PhysRevB.45.7430,spin_dilution_PhysRevB.16.542,spin_dilusion_PhysRevB.45.12548}. In contrast, Aharony et al. \cite{Aharony_PhysRevLett.60.1330} proposed that, in hole-doped cuprates, the doped holes give rise not to simple spin dilution but to spin frustration. The resulting interaction between oxygen holes and copper spins induces an effective ferromagnetic Cu–Cu coupling that competes with the intrinsic antiferromagnetic superexchange. Consequently, even a small amount of hole doping can destabilize the AFM order. These complementary scenarios point out that the doped carriers may fundamentally alter the magnetic behavior.



Recent scanning tunneling microscopy (STM) studies on lightly doped Ca$_2$CuO$_2$Cl$_2$ have further underscored the critical role of the doping potential \cite{ye2023visualizing, yayu_wang_2025}. In the dilute limit, a localized Zhang–Rice singlet has been directly imaged \cite{li2024boundstatesdopedcharge}, demonstrating that doped holes can become trapped by the poorly screened Coulomb potential characteristic of charge-transfer insulators. More importantly, recent advances in angle-resolved photoemission spectroscopy (ARPES) on multilayer cuprate superconductors have provided access to the inner CuO$_2$ planes, which are effectively shielded from disorder. These measurements reveal that an AFM metallic phase with small Fermi pockets can persist up to at least $8\%$ hole doping \cite{Xingjiang_Zhou,clean-cuprate-1,clean-cuprate-2}. Taken together, these observations suggest that the rapid disappearance of AFM order in hole-doped cuprates is likely driven by extrinsic factors. 
Motivated by this perspective, in the following section we analyze how the doping potential impacts magnetic order in cuprates.

\begin{figure}[t]
  \centering
  \includegraphics[width=0.45\textwidth]{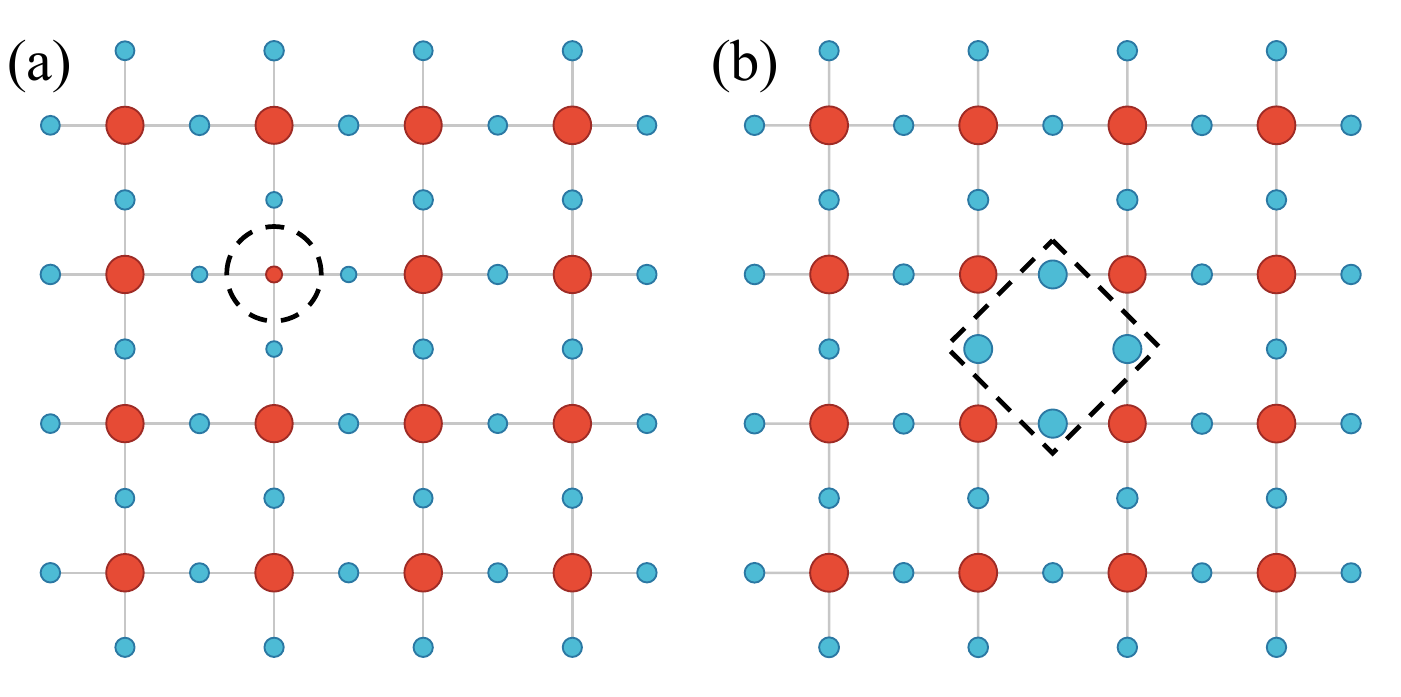}
  \caption{Dopant-induced defect potentials on the CuO$_2$ lattice plane. Red dots denote Cu sites; blue dots denote O sites. Dashed boundaries (circle or square) indicate the set of lattice sites influenced by a single dopant. (a) Electron-doped side: Cu sites within the dashed region experience a local on-site shift $E_{loc}^\text{Cu}$. (b) Hole-doped side: O sites within the dashed region experience $E_{loc}^\text{O}$.}
  \label{fig_defect}
\end{figure}

To mimic dopant–induced local potentials, we augment the Emery Hamiltonian with a defect term~\cite{li2024boundstatesdopedcharge},
\begin{equation}
H_{\text{defect}}=\sum_{j\in D_{\text{Cu}},\sigma}E_{loc}^{\text{Cu}}d_{j\sigma}^{\dagger}d_{j\sigma}+\sum_{j'\in D_{\text{O}},\sigma}E_{loc}^{\text{O}}p_{j'\sigma}^{\dagger}p_{j'\sigma},
\end{equation}
where the parameters $E_{loc}^\text{Cu}$ and $E_{loc}^\text{O}$ represent dopant–induced on-site energy shifts on Cu and O, respectively. $D_{\text{Cu}}$ and $D_{\text{O}}$ denote the sets of Cu and O sites subject to defect potentials.
For electron-doped cuprates, the major effect occurs on the Cu sites and we add a positive $E_{loc}^{\text{Cu}}$ potential to repel further hole occupation. This is equivalent to electron doping, mimicing the effect of $Ce$ or $Zn$, as indicated in Fig.~\ref{fig_defect}(a). 
Motivated by experimental observations in CCOC \cite{ye2023visualizing,yayu_wang_2025}, attractive potentials for holes on oxygen sites within plaquettes are introduced for hole-doped cuprates, as plotted in Fig.~\ref{fig_defect}(b). 
At doping $x$, the total number of defect sites satisfies $\left|D_{\text{Cu}}\right|,\left|D_{\text{O}}\right|\propto\left|x\right|N_{c}$ (with $N_c$ the number of unit cells).
To reduce complexity, we do not perform a disorder average in this study; instead, for each $x$ we select one random defect configuration $D_\text{Cu}$ or $D_\text{O}$ to evaluate AFM-related observables on that realization.

\begin{figure*}[t]
  \centering
  \includegraphics[width=0.98\textwidth]{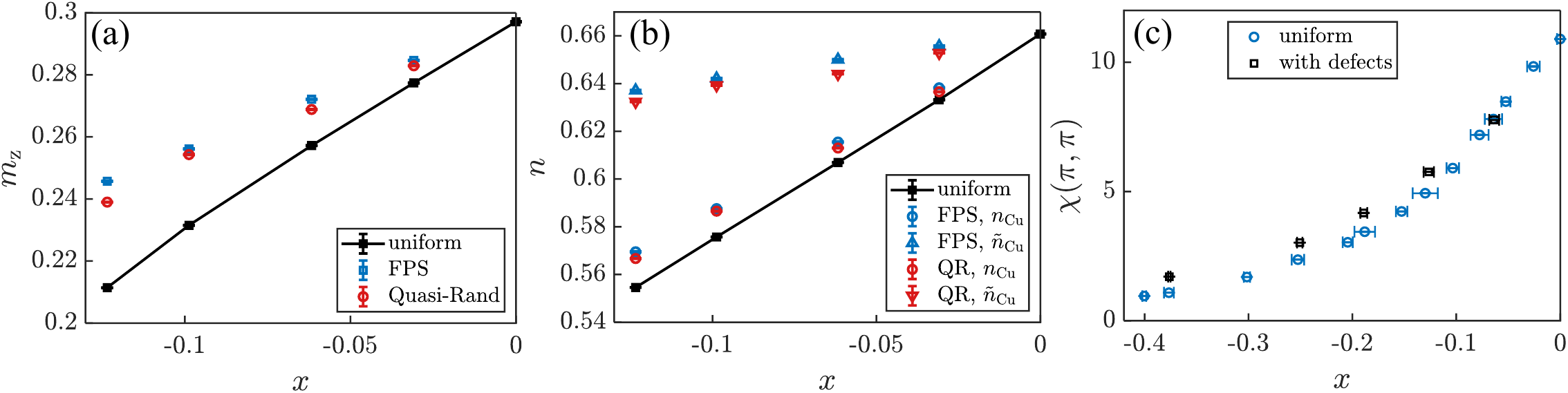}
  \caption{Comparison of AFM order with and without Cu-site defect potentials $E_{loc}^\text{Cu}=2.0$. (a) VMC estimate of the staggered AFM order on $L=18$ clusters. Defect sets are generated by farthest-point sampling (FPS) or quasi-random (QR) sequences (algorithms described in the Appendix). (b) VMC average hole density on Cu sites; circles denote the mean density $n_\text{Cu}$ over all Cu sites; triangles denote the mean density $\tilde{n}_\text{Cu}$ over nondefect Cu sites. (c) DQMC staggered susceptibility $\chi(\pi,\pi)$ for $L=8$, $\beta=8$ where defect positions are chosen by the QR method.}
  \label{fig8}
\end{figure*}

We begin with the electron-doped side, where we introduce Cu-site defect potentials by assigning $|D_\text{Cu}|=|x|N_c$ defect sites and a local potential strength $E_{loc}^\text{Cu}=2.0$.
To sample defect configurations, we employ both farthest-point sampling (FPS) and quasi-random (QR) sequences; technical details are provided in the Appendix. Magnetic properties are computed using VMC and DQMC. For VMC ((L = 18); ansatz details in the Appendix), we optimize the AFM state and evaluate the resulting staggered magnetization.
For DQMC, we calculate the staggered spin susceptibility on (L=8) clusters at ($\beta=8$).
As shown in Fig.~\ref{fig8}(a), the VMC results reveal a clear enhancement of $m_z$ at every electron-doping level.
Although FPS and QR sampling yield slightly different quantitative values, both exhibit the same overall trend of strengthened AFM order under Cu-site defect potentials. This enhancement is independently confirmed by the DQMC susceptibility in Fig.~\ref{fig8}(c).

To further understand this effect, we examine the hole density distribution across Cu sites, shown in Fig.~\ref{fig8}(b) (and Fig.~\ref{fig_s4} in Appendix \ref{Appendix:F}).
The defect potential depletes hole occupation on the defect Cu sites—effectively introducing additional local electron doping.
Meanwhile, the hole density on non-defect Cu sites is only weakly reduced with increasing electron doping, remaining close to the undoped value (Fig.~\ref{fig8}(b), blue and red triangles).
Thus, the non-defect Cu sublattice behaves as if it experiences a reduced effective doping, consistent with a microscopic realization of Cu-site spin dilution. This mechanism naturally enhances AFM robustness, in line with the classic spin-dilution scenario \cite{spin_dilusion_PhysRevB.45.12548}.

\begin{figure}[t]
  \centering
  \includegraphics[width=0.45\textwidth]{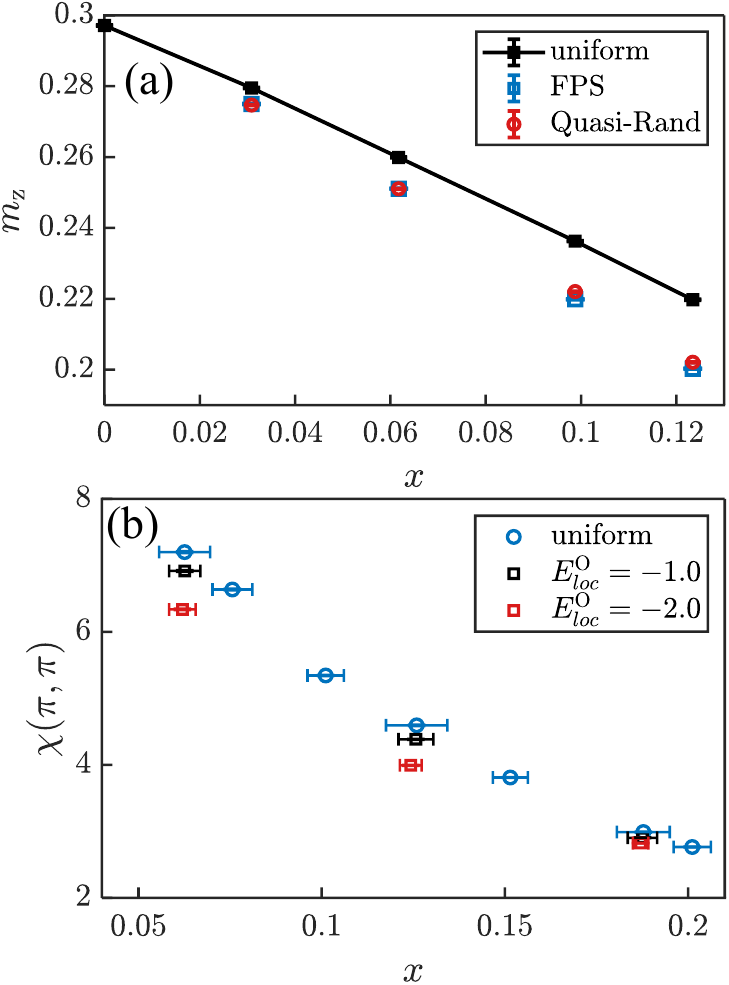}
  \caption{Comparison of AFM order with and without O-site defect potentials $E_{loc}^\text{O}=-1.0$. (a) VMC estimate of the staggered AFM order on $L=18$ clusters. (b) DQMC estimate of the staggered susceptibility $\chi(\pi,\pi)$ for $L=8$, $\beta=8$, where defect positions are chosen by the QR method.}
  \label{fig10}
\end{figure}

Then, we turn to the hole-doped side, where the doping potential is applied on the oxygen sites.
In this case, we set $|D_\text{O}|=4xN_c$ and use a local potential strength $E_{loc}^\text{O}=-1.0$.
Defect configurations are generated using the same FPS and quasi-random protocols as in the Cu-defect analysis. Fig.~\ref{fig10} summarizes the AFM response obtained from VMC and DQMC using the same computational settings as before (VMC: $L=18$; DQMC: $L=8$, $\beta=8$).
Both methods show that O-site defects suppress antiferromagnetic order.
In the zero-temperature VMC results [Fig.~\ref{fig10}(a)], the two defect-sampling strategies yield nearly identical reductions in $m_z$, demonstrating that the suppression is robust and largely insensitive to the detailed defect configuration.
In the finite-temperature DQMC data [Fig.~\ref{fig10}(b)], the suppression is more modest for $E_{loc}^\text{O}=-1.0$ (black squares), likely reflecting thermal effects; this warrants further investigation.
Nevertheless, both methods consistently indicate that oxygen-site defects weaken AFM order on the hole-doped side.

\begin{figure}[t]
  \centering
  \includegraphics[width=0.465\textwidth]{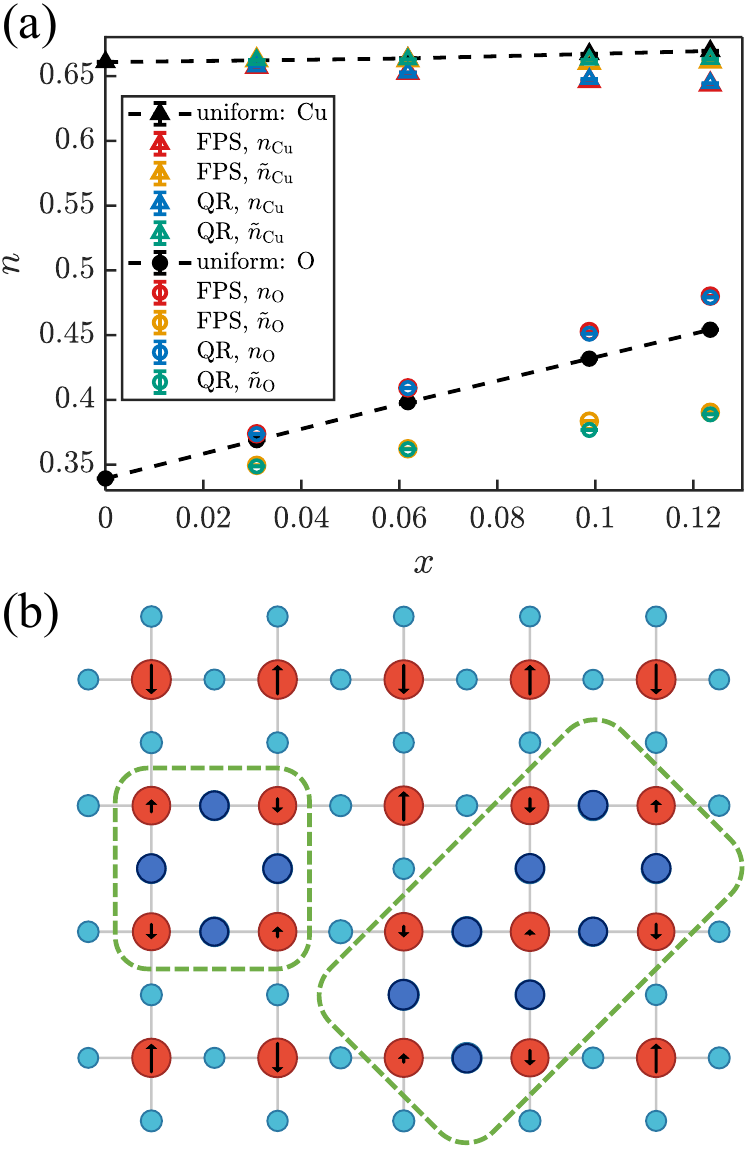}
  \caption{(a) VMC orbital-resolved hole densities. Triangles denote Cu-site hole densities; circles denote O-site hole densities. $n_\text{Cu}$ and $n_\text{O}$ are averages over all Cu and O sites, respectively; $\tilde{n}_\text{Cu}$ and $\tilde{n}_\text{O}$ are averages restricted to nondefect sites (Cu sites not adjacent to the O-defect set and O sites outside the defect set). (b) Partial view of the VMC real-space ground-state charge and spin textures on an $L=18$ lattice with 32 doped holes (x=32/$N_c$). Red dots denote Cu sites; blue dots denote O sites. Arrow length encodes the local magnetic moment. Green dashed outlines indicate the O-defect locations and the main affected region.}
  \label{fig11}
\end{figure}

To probe the origin of this suppression, we examine ground-state charge and spin textures (Fig.~\ref{fig11}). Relative to the uniform (no-defect) case, introducing the O-defect potential slightly reduces the unit-cell–averaged Cu-hole density and increases the total O-hole density; however, the hole density on nondefect oxygen sites, $\tilde{n}_\text{O}$, is markedly reduced, while the hole density on nondefect Cu sites (Cu not bonded to defect O sites), $\tilde{n}_\text{Cu}$, remains essentially unchanged [Fig.~\ref{fig11}(a)]. This indicates a redistribution of charge from other O sites—and from Cu near the defect—onto the defect O sites themselves, which is evident in the charge map of Fig.~\ref{fig11}(b). Concomitantly, the real-space spin texture shows a pronounced reduction of local moments on Cu directly bonded to an O-defect, providing the primary source of the weakened AFM order. A plausible microscopic interpretation is that the defect potential on O, together with the enhanced hole occupancy there, lowers the effective superexchange $J_\text{eff}$ between neighboring Cu spins, thereby reducing local moments around the defect and suppressing AFM order on the hole-doped side.

\section{Summary and Perspective}
\label{sec:summary}

 In summary, we revisit the long-standing question of electron–hole asymmetry of AFM in cuprates, employing the fundamental three-band Emery model.
After establishing a realistic parameter set for La$_2$CuO$_4$ via benchmarks to experimental spin-wave data, we employ a powerful suite of numerical methods, including VMC, DQMC, CP-AFQMC, DMET, and GA, to study the pristine CuO$_2$ plane. Our key finding is that, within this three-band model, the AFM response to doping is almost \textit{symmetric} with respect to electron and hole dopings if we don't consider other possible states such as stripe order \cite{YANAGISAWA2002292,White_PhysRevB.92.205112,PhysRevB.102.214512,PhysRevB.108.205154,Huang2017Science,doi:10.1073/pnas.2408717121}.
This conclusion is robust against a moderate oxygen-site repulsion $U_p$ and against material-specific variations. With parameters appropriate for Nd$_2$CuO$_4$, the electron-doped AFM response closely resembles that of La$_2$CuO$_4$ and no electron–hole asymmetry is observed.

Motivated by recent experiments \cite{yayu_wang_2025,li2024boundstatesdopedcharge,ye2023visualizing}, we also investigate the role of dopant-induced defect potentials. Indeed, a strong asymmetry naturally emerges in this scenario. On the electron-doped side, the Cu-site defects enhance AFM order by restoring neighboring sites toward the robust half-filled magnetic state, which is consistent with a spin-dilution–like mechanism in Heisenberg model~\cite{lco_PhysRevB.45.7430,spin_dilusion_PhysRevB.45.12548,spin_dilution_PhysRevB.16.542}. On the hole-doped side, the O-site defects suppress AFM order by localizing charges around the defected sites and then disrupting the Cu-O-Cu superexchange pathways.
Overall, our work suggests that the defects created by the dopant play a significant role in the electron–hole asymmetry in cuprate AFM magnetism of the correlated CuO$_2$ plane.

Our study highlights a consistent, cross-method picture for the clean Emery model and show extrinsic dopant plays a significant role in the apparent asymmetries reported in some systems. While the present work focuses on the AFM solution of the three-band Emery model, we have not addressed possible competing orders. A natural next step is to treat the intertwined landscape—charge/spin stripes \cite{YANAGISAWA2002292,White_PhysRevB.92.205112,PhysRevB.102.214512,PhysRevB.108.205154,Huang2017Science,doi:10.1073/pnas.2408717121}, superconductivity (including uniform and pair-density-wave states) \cite{PhysRevResearch.2.043259,PhysRevB.107.214504,PhysRevB.103.144514,doi:10.1073/pnas.2106476118}, and related fluctuations—on equal footing with anti-ferromagnetism. On the defect side, our VMC ansatz is intentionally minimal and largely neglects defect–defect interactions; richer variational families are warranted. Recently, the development of neural quantum states (NQS) provide a powerful class of ansatz that substantially enlarge the accessible variational manifold and have shown promise for strongly correlated lattices\cite{Lv2025arXiv,NQS_pfaffian1} and disordered systems\cite{ge2025quencheddisorder}. Deploying NQS for the defected Emery model should yield more accurate energetics and correlation profiles, and enable to resolve defect-defect interactions. Finally, the microscopic mechanism by which O-site defects suppress hole-side AFM—beyond the phenomenology reported here—remains unresolved and targeted studies that quantify local charge transfer and the accompanying reduction of effective superexchange near O defects will be essential.

\section{Acknowledgement}
We acknowledge the support by the Ministry of Science and Technology  (Grant No. 2022YFA1403900, No. 2022YFA1405400), the National Natural Science Foundation of China (Grant No. NSFC-11888101, No. NSFC-12174428, No. NSFC-11920101005, No. NSFC-12274290, NSFC-12474146, NSCF-12347107, No. NSFC-12274290, and No. 12522406), the New Cornerstone Investigator Program, the Chinese Academy of Sciences Project for Young Scientists in Basic Research (2022YSBR-048), the Innovation Program for Quantum Science and Technology (2021ZD0301902) and the Beijing Natural Science Foundation (No. JR25007).
HKJ acknowledges the support from the start-up funding from ShanghaiTech University.

\bibliography{reference}


\newpage
\clearpage

\appendix

\section{Different conventions for the Emery model}

\begin{figure}[t]
  \centering
  \includegraphics[width=0.46\textwidth]{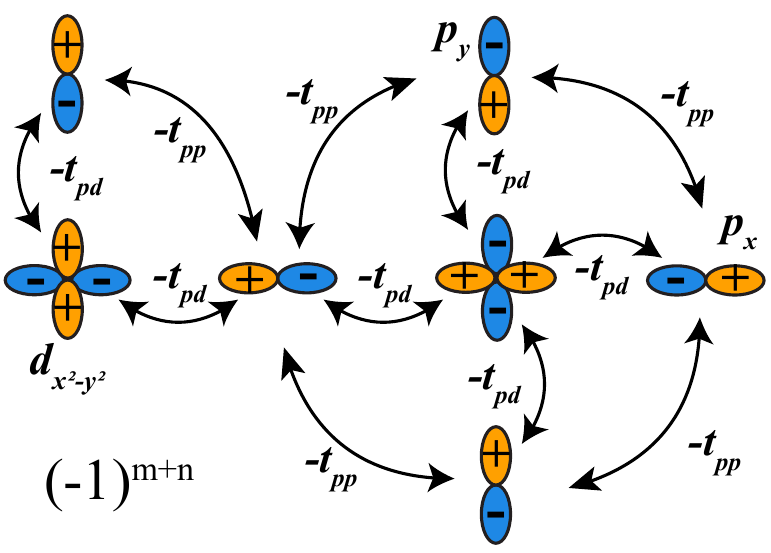}
  \caption{Hopping pattern of the three-band Emery model in the Emery–Reiter convention. The Cu site in each unit cell is located at $(m,n)$, with the two O sites at $(m+\tfrac{1}{2},n)$ and $(m,n+\tfrac{1}{2})$. A staggered phase factor $(-1)^{m+n}$ is applied to the orbital basis, making all nearest-neighbor Cu–O and O–O hopping amplitudes homogeneous in sign.}
  \label{Emery2}
\end{figure}

In the main text we use a particular hole representation of the three-band Emery model. An alternative convention, due to Emery and Reiter~\cite{Emery_PhysRevB.38.4547}, is often convenient for numerical calculations. For a unit cell at integer coordinates \((m,n)\) (with a Cu site at \((m,n)\) and two O sites at \((m+\tfrac{1}{2},n)\) and \((m,n+\tfrac{1}{2})\)), they flip all the $p_y$ orbitals with a negative sign and introduce an additional staggered phase factor \((-1)^{m+n}\) on the orbital basis. This gauge transformation renders the hopping amplitudes spatially homogeneous in sign. The corresponding hopping pattern in this convention is shown in Fig.~\ref{Emery2}.

\section{VMC}
The conventional VMC optimization is performed using the stochastic reconfiguration (SR) method. A key step in this process is computing the variational energy and its gradient with respect to the set of variational parameters, denoted $\{\theta\}$.
The variational energy $E(\theta)$ is the expectation value of the Hamiltonian $H$, calculated as:
\begin{equation}
    E(\theta) = \frac{\langle \Psi(\theta) |H| \Psi(\theta) \rangle}{\langle \Psi(\theta) \mid \Psi(\theta) \rangle} = \frac{\sum_C |\Psi(C; \theta)|^2 E_{\mathrm{loc}}(C)}{\sum_C |\Psi(C; \theta)|^2},
    \label{eq:A2}
\end{equation}
where $E_{\mathrm{loc}}(C)$ is denoted as the local energy for a given configuration $|C\rangle$:
\begin{equation}
    E_{\mathrm{loc}}(C) = \sum_{C'} \frac{\langle C|H|C' \rangle \Psi(C'; \theta)}{\Psi(C; \theta)}.
\end{equation}
To minimize the energy, we require its gradient, which is given by:
\begin{equation*}
    \vec{f}\equiv-\nabla_{\theta} E(\theta) = -2 \Re[\left( \langle \nabla_{\theta} (\ln\Psi) E_{\mathrm{loc}} \rangle + \langle E_{\mathrm{loc}} \rangle \langle \nabla_{\theta} (\ln\Psi) \rangle \right)].
\end{equation*}
The expectation values $\langle \cdot \rangle$ are computed via standard MC sampling over the probability distribution $|\Psi(C; \theta)|^2$.

The SR method accelerates convergence by approximating the Hessian of the energy with a positive-definite Hermitian matrix $S$, also known as the quantum geometric tensor. This avoids the high computational cost of the true Hessian while providing an improved search direction. The elements of $S$ are given by the covariance matrix of the logarithmic wave function derivatives:
\begin{equation}
    S_{ij} = \langle (\nabla_{\theta_i} \ln\Psi) (\nabla_{\theta_j} \ln\Psi) \rangle - \langle \nabla_{\theta_i} \ln\Psi \rangle \langle \nabla_{\theta_j} \ln\Psi \rangle.
    \label{eq:A5}
\end{equation}
With the matrix $S$, the parameter update step $\Delta\vec{\theta}$ is determined by solving the following linear system:
\begin{equation}
    \Delta\vec{\theta} = \frac{\Delta\tau}{2} S^{-1} \vec{f}.
    \label{eq:A6}
\end{equation}

As mentioned in the main text, we use two types of variational wave functions. The first one is from the open source software package mVMC,
\begin{equation}
\left|\Psi\right\rangle =\mathcal{P}_{G}\mathcal{P}_{J}\mathcal{P}_{S}\left|\phi_{pair}\right\rangle ,
\end{equation}
where the pair-product wave function $\left|\phi_{pair}\right\rangle$ takes the form of
\begin{equation}
\left|\phi_{pair}\right\rangle =\left(\sum_{i,j=1}^{N_{c}}\sum_{\mu_{1},\,\mu_{2}\in\{d,p_{x},p_{y}\}}f_{ij}^{\mu_{1}\mu_{2}}c_{i\mu_{1}\uparrow}^{\dagger}c_{j\mu_{2}\downarrow}^{\dagger}\right)^{N_{e}/2}\left|0\right\rangle .
\end{equation}
Here $f_{ij}^{\mu_{1}\mu_{2}}$ is a variational parameter, $N_e$ is number of holes (since we are working in the hole representation throughout this paper), and $\left|0\right\rangle $ is a vacuum. In the calculations in the main text, we take $2\times2$ sublattice structures to consider off-site correlations. $\mathcal{P}_S$ is the spin quantum-number projector whose explicit form can be found in Refs.~\cite{vmc_MISAWA2019,vmc_Imada2008}. The Gutzwiller factor controls the number of the doubly occupied sites through the variational parameters $g_i^\mu$ defined at each site as below,
\begin{equation}
\mathcal{P}_{G}=\exp\left(-\sum_{i,\mu}g_{i}^{\mu}n_{i\mu\uparrow}n_{i\mu\downarrow}\right).
\end{equation}
The Jastrow factor introduces long-range charge–charge correlations, which is defined as,
\begin{equation}
\mathcal{P}_{J}=\exp\left(-\frac{1}{2}\sum_{ij\mu_{1}\mu_{2}}v_{ij}^{\mu_{1}\mu_{2}}n_{i\mu_{1}}n_{j\mu_{2}}\right).
\end{equation}
Here, we set $v_{ij}^{\mu_1\mu_2}=0$ for $i=j$ and $\mu_1=\mu_2$ since the on-site correlation is already introduced by $\mathcal{P}_G$. We also assume the correlations factors take the same $2\times2$ sublattice structure in the calculations.

In the mVMC calculations, spin quantum-number projection $\mathcal{P}_S$ is applied only for systems with linear size $L\leq12$ to reduce computational cost. For both variational optimization and observable evaluation, we use  $N_\text{MC}=7.6\times10^4\,\text{--}\,9.6\times10^4$ Monte Carlo samples. The SR updates employ a discrete time step $\Delta\tau=0.01\text{--}0.02$.

For the conventional VMC, the variational wave functions are
\begin{equation}
\left|\Psi\right\rangle =\mathcal{P}_{G}\left|\phi_{mf}\right\rangle ,
\end{equation}
where $\left|\phi_{mf}\right\rangle $ is generated through a mean-field Hamiltonian,
\begin{align}
H_{\text{mf}} & =\sum_{\left\langle il\right\rangle ,\sigma}\chi_{01}\left(d_{i\sigma}^{\dagger}p_{l\sigma}+h.c.\right)-\sum_{\left\langle ll'\right\rangle ,\sigma}\chi_{11}\left(p_{l\sigma}^{\dagger}p_{l'\sigma}+h.c.\right)\nonumber \\
 & +\sum_{\left\langle ij\right\rangle ,\sigma}\chi_{00}\left(d_{i\sigma}^{\dagger}d_{j\sigma}+h.c.\right)+\epsilon_{p}\sum_{l,\sigma}p_{l\sigma}^{\dagger}p_{l\sigma}\nonumber \\
 & +m_{z,00}\sum_{i}(-1)^{x+y}\left(d_{i\uparrow}^{\dagger}d_{i\uparrow}-d_{i\downarrow}^{\dagger}d_{i\downarrow}\right)\nonumber \\
 & +m_{z,11}\sum_{l}(-1)^{x+y}\left(p_{l\uparrow}^{\dagger}p_{l\uparrow}-p_{l\downarrow}^{\dagger}p_{l\downarrow}\right).
\end{align}
Here $(x,y)$ is the 2D index of the $i$-th unit cell and the last two terms encode AFM order on the Cu and O sublattices, respectively. In our conventional VMC setup we restrict the correlator to on-site terms, consistent with the dominance of local Hubbard interactions in this model. Specifically, the Gutzwiller projector is
\begin{equation}\label{eqn:vmc_gutz}
\mathcal{P}_{G}=\exp\left[-g_{d}\sum_{i}n_{id\uparrow}n_{id\downarrow}-g_{p}\sum_{i}\left(n_{ip_{x}\uparrow}n_{ip_{x}\downarrow}+n_{ip_{y}\uparrow}n_{ip_{y}\downarrow}\right)\right],
\end{equation}
where the O-orbital factor $g_p$ is included only when $U_{p}\neq0$. In the VMC calculations, we fix the $\chi_{01}\equiv1.0$ (as a choice of scale) and treat all other mean-field parameters and projector strengths as variational. For both variational optimization and observable evaluation, we use  $N_\text{MC}=5.8\times10^4\,\text{--}\,9.6\times10^4$ Monte Carlo samples. The SR updates employ a discrete time step $\Delta\tau=0.01\text{--}0.04$.

\section{DQMC}

\begin{figure}[t]
  \centering
  \includegraphics[width=0.47\textwidth]{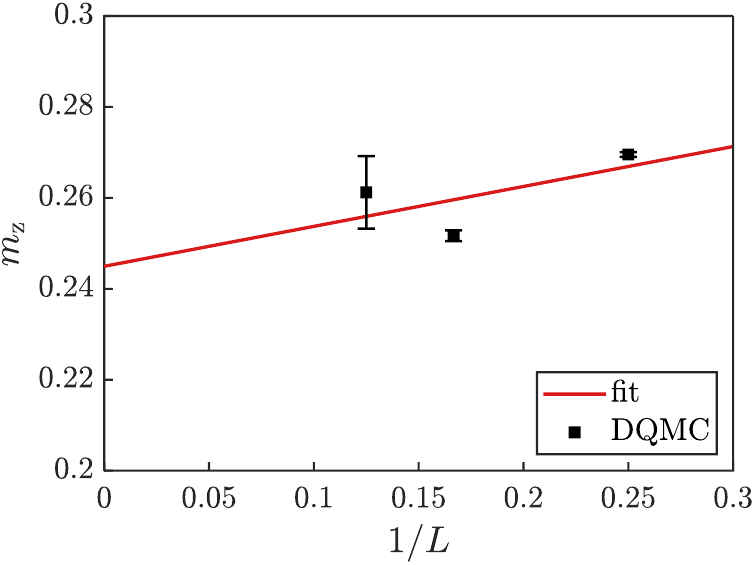}
  \caption{Undoped ($x=0$) magnetism in the three-band Hubbard model with parameters in \eqref{parameter} esitimated by DQMC. The magnetic moment is extracted from spin correlation factor $S(\pi,\pi)$ for each lattice size $L$ at inverse temperature $\beta=2L$; a linear $1/L$ fit is used to obtain the thermodynamic limit estimate.}
  \label{dqmc_appendix}
\end{figure}

 In our DQMC calculations, we study the unbiased finite-temperature properties by evaluating the expectation values of observables under grand canonical ensemble average.  The inverse temperature is discretized as $\beta=L\Delta\tau$ and the partition function is then expressed as:
\begin{equation}
Z= {\rm Tr}\left[e^{-\beta (\hat{H} - \mu \hat{N})}\right] = {\rm Tr}\left[\prod_{l=1}^{L_{\tau}}e^{-\Delta_{\tau} (\hat{H} - \mu \hat{N})}\right]
\label{pf}
\end{equation}
and the imaginary-time evolution operator is factorized using the symmetric second-order Trotter--Suzuki decomposition:
\begin{equation}
e^{-\Delta\tau (H-\mu N)}
\simeq
e^{-\frac{\Delta\tau}{2}H_t}\,
e^{-\Delta\tau H_V}\,
e^{-\frac{\Delta\tau}{2}H_t}.
\end{equation}
Here $H-\mu N$ is rewritten as the sum of a kinetic and a potential term,
\begin{equation}
H-\mu N = H_t + H_V,
\end{equation}
with
\begin{align}
H_t &= 
\sum_{i\alpha,j\alpha',\sigma}
t_{i\alpha,j\alpha'}
\left( c^{\dagger}_{i\alpha\sigma} c_{j\alpha'\sigma} + \mathrm{h.c.} \right), \\
H_V &= 
\sum_{i,\alpha}
U_{i\alpha} n_{i\alpha\uparrow} n_{i\alpha\downarrow}
+\sum_{i,\alpha} (\epsilon_{i\alpha}-\mu) n_{i\alpha},
\end{align}
where $i$ and $j$ label unit cells and $\alpha,\alpha'$ denote orbital indices.

The onsite Hubbard interaction is decoupled on each time slice $l$ in the $S^z$ channel using the discrete Hubbard--Stratonovich transformation:
\begin{equation}
e^{-\Delta\tau U_{i\alpha} n_{\uparrow} n_{\downarrow}}
=
\frac{1}{2}\sum_{s=\pm1}
\exp\!\bigg[
\lambda_{i\alpha} s (n_{\uparrow}-n_{\downarrow})
-\frac{\Delta\tau U_{i\alpha}}{2}(n_{\uparrow}+n_{\downarrow})
\bigg],
\end{equation}
with 
\begin{equation}
\lambda_{i\alpha}
=
\mathrm{acosh}\!\left( e^{\Delta\tau U_{i\alpha}/2} \right).
\end{equation}
After the Hubbard-Stratonovich transformation, the fermion interaction is decoupled to the bilinear fermion operator coupled to the space-time dependent auxiliary field. This procedure allows the fermionic degrees of freedom to be integrated out analytically. Consequently, the partition function in Eq.~\eqref{pf} is reformulated as a sum over all possible configurations of the auxiliary field. The weight of each configuration, $W_s$, is given by the determinant of the resulting fermion matrix. The summation can be evaluated stochastically via Monte Carlo sampling. For the update schemes and measurement techniques used in DQMC, we refer the reader to established reviews\cite{AssaadReview,Li2019review}.

Our DQMC simulations are performed on systems up to $N_s = 8 \times 8$ sites and at temperatures down to $T = 1/8$ and $T = 1/10$. For each set of parameters, we adjust the chemical potential $\mu$ to fix the target doping level. 
The accessible parameter space of system size and temperature is ultimately dependent on the severity of the fermion sign problem, which varies with doping. To ensure robust statistics, our results are averaged over 96--384 independent Markov chains. Each chain consists of 25 warm-up (equilibration) sweeps followed by a minimum of 250--500 measurement sweeps. To maintain statistical precision in parameter regimes where the sign problem is more severe, the number of measurement sweeps is increased accordingly. 

To study the magnetic excitations compared with experiments, We perform stochastic analytic continuation to obtain the dynamical spin structure factor from the imaginary-time spin correlation function \(G(\mathbf{q},\tau) = \langle S^{z}({\mathbf{q}},\tau) S^{z}({-\mathbf{q}},0)\rangle\) evaluated in the DQMC simulation:
\begin{equation}
G(\mathbf{q},\tau)
= \int_{0}^{\infty} \frac{d\omega}{\pi}
S^{z}(\mathbf{q},\omega)
\left( e^{-\tau\omega} + e^{-(\beta-\tau)\omega} \right).
\label{eq:Gtau}
\end{equation}
We employ a rectangular lattice of size $16 \times 4$ to enhance momentum resolution along the $x$ direction. To obtain a reliable real-frequency spectrum from Stochastic Analytic Continuation, we first generate high-precision imaginary-time correlation functions using a smaller time step $\Delta\tau = 0.05$. This is achieved through extensive Monte Carlo sampling, averaging over 384 independent Markov chains, each evolved for over 2000 sweeps to minimize statistical errors.  For each $S^{zz}(q_x, q_y=\pi, \omega)$, the spin-excitation dispersion $\omega(q_x, q_y=\pi)$ is extracted by performing local Gaussian fits to the spectral peaks.

To determine the ground-state staggered magnetization $m_z$ with DQMC, we compute the spin structure factor $S(\mathbf{Q})$ for various system sizes $L$. We fix an inverse temperature $\beta=2L$ that scales with the system size. We perform a finite-size scaling extrapolation of $\sqrt{S(\mathbf{Q})}$ versus $1/L$ through a linear fit, with the intercept yielding the staggered magnetization at zero temperature and thermodynamic limit. The results are shown in Fig.~\ref{dqmc_appendix}, which gives rise to $m_z = 0.245 \pm 0.021$.

\section{CP-AFQMC}

In the CP-AFQMC calculation we utilize the spin decomposition of Hubbard-Stratonovich transformation in $S_z$ channel:
\begin{equation}
e^{-\Delta \tau U n_{i \uparrow} n_{i \downarrow}}=e^{-\Delta \tau U\left(n_{i \uparrow}+n_{i \downarrow}\right) / 2} \sum_{x_i= \pm 1} \frac{1}{2} e^{\gamma x_i\left(n_{i \uparrow}-n_{i \downarrow}\right)},
\end{equation}
where $x_i$ is the auxiliary field and the constant $\gamma$ is determined by 
\begin{equation}
\cosh (\gamma)=e^{\Delta \tau U / 2}.
\end{equation}
We choose $\Delta \tau = 0.05$ as the time step in trotter decomposition which ensures the trotter error is negligible in the results. The typical total number of the walkers $N_w$ is about $10000$ and the convergence of the results regarding $N_w$ is also checked.

As introduced in the main text, the self-consistent optimization of the trial wave-function in CP-AFQMC can be carried out in two different ways, i.e., coupling to a mean-field calculation and coupling to natural orbitals. The two approaches give consistent results.

The coupled mean-field Hamiltonian has the following form \cite{PhysRevB.94.235119}: 
\begin{equation}
{H_{mf}} = K  + U_{\text{eff}} \sum\limits_{i \in {\rm{Cu}}} {\left\langle {{n_{i \uparrow }}} \right\rangle {n_{i \downarrow }}}  + U_{\text{eff}} \sum\limits_{i \in {\rm{Cu}}} {\left\langle {{n_{i \downarrow }}} \right\rangle {n_{i \uparrow }}} 
\label{eq:CP-MF}
\end{equation}
where $K$ denotes the non-interacting terms in the 3-band Hamiltonian, $\left\langle {{n_{i \uparrow }}} \right\rangle$ and $\left\langle {{n_{i \downarrow }}} \right\rangle$ are CP-AFQMC results at the previous step and $U_{\text{eff}}$ is the variational parameter to be optimized so that the mean-field solution of $\langle n_i\rangle$ match the input values from QMC the best. Then the solution (a Slater determinant) of the mean-field Hamiltonian with the optimized $U_{\text{eff}}$ will be used for the next step QMC calculation.

The reduced density matrix approach proceeds as follows \cite{PhysRevB.107.235124}: the one-body Green's function \(G_{ij}\), calculated via the mixed estimator \(G_{ij} = \frac{\langle \psi_T | c_i^{\dagger} c_j | \psi \rangle}{\langle \psi_T | \psi \rangle}\), is diagonalized in the form of \(U V U^\dagger\). The natural orbitals corresponding to the largest \(N_e\) eigenvalues (occupations) are then selected to form the new trial wave function. 


In the hole-doped case where the self-consistent CP-AFQMC calculation doesn't give AFM state but stripe state \cite{PhysRevB.102.214512}, we can enforce an AFM state in the CP-AFQMC calculation. We optimize a AFM-type mean-field trial wave-function by minimizing the CP-AFQMC energy against the AFM order parameter $\Delta$ in the mean-filed Hamiltonian:
\begin{equation}
H_{\text{AF-MF}}=K+\Delta \sum_{i \in \mathrm{Cu}}(-1)^{i} S_{i}^z
\end{equation}
with $K$ the non-interaction terms in the three band Hamiltonian.

\section{DMET}
\subsection{DMET method details}
\begin{figure}[htbp]
  \centering
  \includegraphics[width=0.4\textwidth]{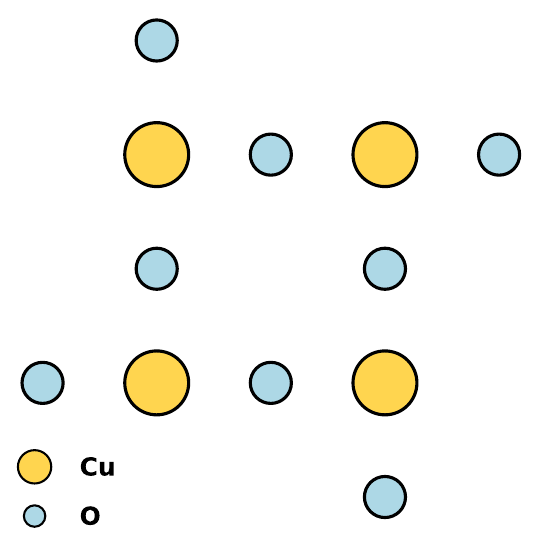} 
  \caption{Impurity unit cell preserving $C_4$ symmetry}
  \label{fig:unit_cell}
\end{figure}

As described in the main text, to preserve $C_4$ symmetry, the impurity unit cell can be chosen as the Fig.~\ref{fig:unit_cell}.
We describe the low-level (mean-field) system by Slater determinants.
And introduce an auxiliary quadratic operator that generates the mean-field solution,
\begin{equation}
\hat{h}
= \hat{h}_{0} + \hat{u}
= \hat{h}_{0} + \sum_{i j \sigma} v_{i j}^{\sigma}\, c_{i\sigma}^{\dagger} c_{j\sigma},
\label{eq:correction_potential}
\end{equation}
As stated in the main text, our chosen correlation potential preserves spin U(1) symmetry along the $z$ axis. In general, the form of the correlation potential can be tailored to the problem under study; for example, one may include additional terms such as superconducting pairing fields \cite{DMET_guide}.
The correlation potential are updated self-consistently within DMET, playing a role analogous to the hybridization function in DMFT \cite{DMFT_1996RMP}.

To construct the embedding problem, we exploit the fact that the low-level Hamiltonian is easy to solve.
We partition the original lattice into a local fragment \(A\) and the remaining environment \(B\).
In the mean-field state, the one-particle density matrix has the block form
\begin{equation}
\rho=\begin{pmatrix}\rho_{A} & \rho_{AB}\\
\rho_{BA} & \rho_{B}
\end{pmatrix}.
\end{equation}
In general, \(A\) and \(B\) are entangled.
One can identify up to \(n_A\) bath orbitals (where \(n_A\) is the number of fragment orbitals) in two equivalent ways \cite{DMET_PhysRevLett.109.186404,DMET_guide}:

(i) By diagonalizing the environment block \(\Gamma_{B}\) and selecting the eigenvectors with eigenvalues strictly between \(0\) and \(1\) (eigenvalues \(0\) or \(1\) correspond to unentangled core/virtual orbitals);

(ii) For a Slater-determinant reference, by performing a singular value decomposition of the fragment–environment block $\rho_{AB}=L\Sigma R^{\dagger}$, and taking the columns of \(R\) associated with the nonzero singular values \(0 < \Sigma_{kk} < 1\) as the bath orbitals in the environment space.
The number of such bath orbitals equals \(\mathrm{rank}(\rho_{A B}) \le n_A\).
The fragment orbitals together with these bath orbitals define the active space \(A_{\mathrm{act}} = A \cup B_{\mathrm{bath}}\).

Let \(\hat{\mathcal{P}}_{\mathrm{emb}}\) denote the projector onto the active space.
Projecting the three-band Hamiltonian onto this space yields the high-level embedding Hamiltonian
\begin{equation}
\hat{H}^{\mathrm{emb}} = \hat{\mathcal{P}}_{\mathrm{emb}}\, \hat{H}\, \hat{\mathcal{P}}_{\mathrm{emb}}.
\label{eq:embedding_Ham}
\end{equation}
 Solving these embedding problems typically dominates the computational cost in the DMET self-consistency cycle. Each embedding problem can be solved using exact diagonalization (ED) \cite{ED}, the density-matrix renormalization group (DMRG) \cite{White_DMRG1,White_DMRG2,Garnet_DMRG}, or coupled cluster singles and doubles (CCSD) \cite{CCSD_method}, which is primarily employed in this work.
We determine the next correlation potential \(\hat{u}\) (with matrix elements \(v_{ij}^{\sigma}\)) by least-squares (LS) minimization of the mismatch between the high-level and low-level one-particle density matrices, either on the fragment block or over the entire lattice:
\begin{equation}
f(u)\;=\;\sum_{\sigma}\,\big\|\rho_{\sigma}^{\mathrm{HL}}-\rho_{\sigma}^{\mathrm{LL}}(u)\big\|_{F}^{2}.
\end{equation}
Here \(\rho^{\mathrm{LL}}(u)\) is computed from \(\hat{h}=\hat{h}_{0}+\hat{u}\), and \(\|\cdot\|_{F}\) denotes the Frobenius norm.
We then rebuild the bath from new correlation potential $\hat{u}^{new}$, reproject \(\hat{H}\) to obtain \(\hat{H}^{\mathrm{emb}}\), solve the high-level problem, and iterate until convergence, e.g.,
\begin{equation}
\big\|\rho^{\mathrm{HL}}-\rho^{\mathrm{LL}}\big\|_{F}<\varepsilon_{\Gamma}\quad\text{and}\quad\|v^{(k+1)}-v^{(k)}\|_{F}<\varepsilon_{v}.
\end{equation}
We use two convergence criteria for the DMET self-consistency cycle: the change in the ground-state energy between successive iterations is below $1\times10^{-5}$, and $\varepsilon_{v} < 1\times10^{-4}$.

DMET results show that the moments are predominantly on Cu sites. We therefore define the magnetic order parameter as the average magnitude of the four Cu moments in the impurity cluster:
\begin{equation}
m_{\mathrm{Cu}} \equiv \frac{1}{4}\sum_{i=1}^{4} \big|m_i^{d}\big|.
\end{equation}

\subsection{Comparison of Different Solvers}
In addition to using CCSD as the impurity solver, we also employ DMRG. Keeping all model parameters unchanged, the DMRG solver uses a bond dimension of $M=1200$. The convergence criteria are slightly relaxed due to the higher computational cost. The results agree very well with those from CCSD in absolute value. The table below compares the two at several representative doping points.
\begin{table}[htbp]
\centering
\setlength{\tabcolsep}{6pt}
\renewcommand{\arraystretch}{1.2}
\begin{tabular}{c|cc|cc}
\hline
 & \multicolumn{2}{c|}{CCSD solver} & \multicolumn{2}{c}{DMRG solver} \\
\cline{2-5}
$x$ & $m_{\mathrm{Cu}}$ & $n_{\mathrm{Cu}}$
            & $m_{\mathrm{Cu}}$ & $n_{\mathrm{Cu}}$ \\
\hline
-0.1 & 0.21959(8) & 0.575142(6)  & 0.21260(54) & 0.574(1) \\
0.0 & 0.2856(1) & 0.660452(6)  & 0.2945(20) & 0.6682(38) \\
0.1 & 0.22493(1) & 0.670256(0)  & 0.21481(13) & 0.6730(4) \\
0.2 & 0.16760(4) & 0.686370(0)  & 0.15062(31) & 0.6913(10)\\
\hline
\end{tabular}\caption{Comparison of results from different impurity solvers}
\end{table}

\begin{figure}[t]
  \centering
  \includegraphics[width=0.48\textwidth]{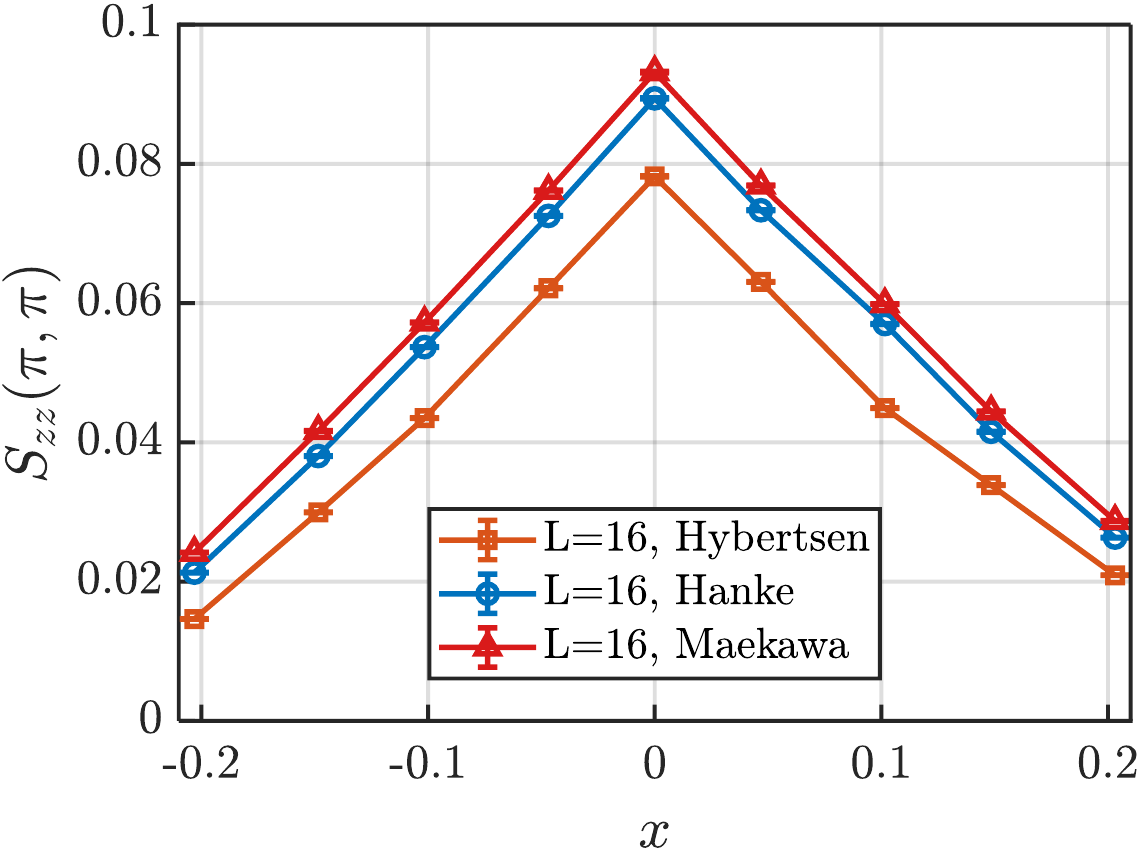}
  \caption{Spin correlation as a function of doping for some other three-band Hubbard models in Table.~\ref{otherparams}}
  \label{fig_s1}
\end{figure}

\section{Parameters in other References}
In addition, we have performed VMC calculations using several alternative three-band Emery parameter sets reported in the literature (listed in Table.~\ref{otherparams}). For each set, we computed the doping dependence of the staggered spin correlation functions in the same manner as in the main text. As shown in Fig.~\ref{fig_s1}, in all cases, the evolution of AFM order with $x$ is qualitatively identical to our La$_2$CuO$_4$-based results, and no pronounced electron–hole asymmetry is resolved within our numerical accuracy. This corroborates that the main conclusions are robust against reasonable variations of the underlying model parameters.
\begin{table}[htbp]
\begin{center}
\begin{tabular}{ |c| c c c c| } 
 \hline\hline
 Model & $t_{pd}$ & $t_{pp}$ & $\Delta_{pd}$ & $U_d$ \\ 
 \hline 
 Hybertsen \cite{Hybertsen_PhysRevB.39.9028} & 1.3 & 0.65 & 3.6 & 10.5 \\ 
 \hline
 Hanke \cite{Hanke_parameter} & 1.0 & 0.5 & 3.0 & 8.0 \\ 
 \hline
 Maekawa \cite{Maekawa_JPSJ.59.1760} &  1.13 & 0.49 &  3.3 &  8.5 \\
 \hline\hline
\end{tabular}\caption{Three-band Emery model parameter sets adopted in the literature and used for additional VMC benchmarks in this work.}\label{otherparams}
\end{center}
\end{table}

\section{Doping potential}\label{Appendix:F}
\subsection{Defect positions}
\begin{figure}[t]
  \centering
  \includegraphics[width=0.48\textwidth]{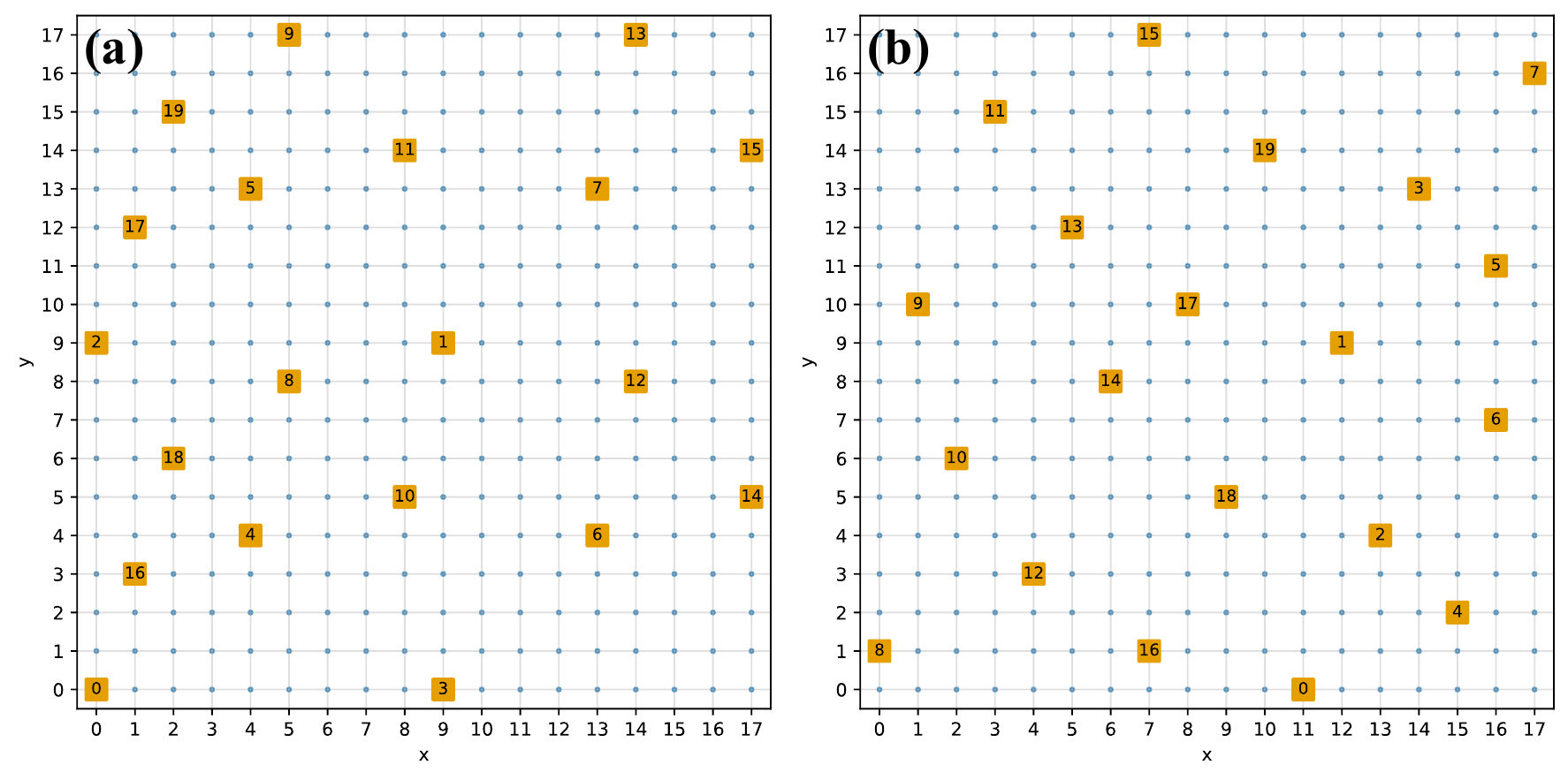}
  \caption{Defect points (orange squares). (a) Selected by farthest-point sampling (FPS). (b) Selected by a quasi-random (QR) sequence.}
  \label{figF1}
\end{figure}

As noted in the main text, we select the set of defect points $D_{\text{pts}}\subset\left\{ 0,...,L_{x}-1\right\} \times\left\{ 0,...,L_{y}-1\right\} $ on a two-dimensional torus using two complementary schemes: farthest-point sampling (FPS) and quasi-random (QR) sequences. Periodic separations between two points $p=(x,y)$ and $q=(u,v)$ are measured with the torus metric
\begin{equation}
d^{2}(p,q)\!\equiv\!\min\left(|x\!-\!u|,L_{x}\!-\!|x\!-\!u|\right)^{2}+\min\left(|y\!-\!v|,L_{y}\!-\!|y\!-\!v|\right)^{2}.
\end{equation}
FPS proceeds via a farthest-first traversal that greedily maximizes the minimal distance to the already selected points, producing a blue-noise–like, well-dispersed set of $k$ defects (cf. González’s farthest-first algorithm for the $k$-center problem \cite{GONZALEZ1985293}). For QR sampling, we generate $k$ points from a low-discrepancy Halton sequence mapped to the torus, with an optional Cranley–Patterson random shift to decorrelate from the lattice \cite{Halton1960,QR2}. Duplicates after quantization are skipped until $k$ unique points are obtained. Fig.~\ref{figF1} illustrates 20 defect points chosen by FPS and by QR on an $18\times18$ torus.
Once the defect points are selected, the defect sites for Cu-site and O-site can be determined according to Fig.~\ref{fig_defect}.

\subsection{Variational ansatz}
To describe the ground state in the presence of Cu-defect, the mean-field wave function $\left|\phi_{mf}\right\rangle $ is generated from the following mean-field Hamiltonian,
\begin{align}
H_{\text{mf}}^{\text{Cu-defect}} & =\sum_{\left\langle il\right\rangle ,\sigma}\chi_{01}\left(d_{i\sigma}^{\dagger}p_{l\sigma}+h.c.\right)-\sum_{\left\langle ll'\right\rangle ,\sigma}\chi_{11}\left(p_{l\sigma}^{\dagger}p_{l'\sigma}+h.c.\right)\nonumber \\
 & +\sum_{\left\langle ij\right\rangle ,\sigma}\chi_{00}\left(d_{i\sigma}^{\dagger}d_{j\sigma}+h.c.\right)+\epsilon_{p}\sum_{l,\sigma}p_{l\sigma}^{\dagger}p_{l\sigma}\nonumber \\
 & +\sum_{i,\sigma}\epsilon_{i}d_{i\sigma}^{\dagger}d_{i\sigma}+\sum_{i}m_{Cu,i}(-1)^{x+y}\left(d_{i\uparrow}^{\dagger}d_{i\uparrow}-d_{i\downarrow}^{\dagger}d_{i\downarrow}\right)\nonumber \\
 & +m_{z,11}\sum_{l}(-1)^{x+y}\left(p_{l\uparrow}^{\dagger}p_{l\uparrow}-p_{l\downarrow}^{\dagger}p_{l\downarrow}\right),
\end{align}
where $\epsilon_i$ and $m_{Cu,i}$ are given by
\begin{equation}
\epsilon_{i}=\begin{cases}
\epsilon_{d0} & \text{if}\;i\in D_{\text{Cu}}\\
\epsilon_{d1} & \text{if}\;i\in D_{\text{Cu}}^{(1)}\\
\epsilon_{d2} & \text{if}\;i\in D_{\text{Cu}}^{(2)}\\
0 & \text{otherwise}
\end{cases},\;m_{\text{Cu},i}=\begin{cases}
m_{z,00}^{(0)} & \text{if}\;i\in D_{\text{Cu}}\\
m_{z,00}^{(1)} & \text{if}\;i\in D_{\text{Cu}}^{(1)}\\
m_{z,00}^{(2)} & \text{if}\;i\in D_{\text{Cu}}^{(2)}\\
m_{z,00} & \text{otherwise}.
\end{cases}
\end{equation}
Here $D_\text{Cu}^{(1)}$ denotes the set of Cu sites that are nearest neighbors of the defect set $D_\text{Cu}$, and $D_\text{Cu}^{(2)}$ denotes the next-nearest neighbors.

For the O-defect case, we define \(D_{\mathrm O}\) as the set of defect oxygen sites and \(\tilde D_{\mathrm{Cu}}\) as the set of Cu sites directly bonded to \(D_{\mathrm O}\); we further define the bond sets \(B_{\mathrm{Cu}\text{–}\mathrm O}\), \(B_{\mathrm O\text{–}\mathrm O}\), and \(B_{\mathrm{Cu}\text{–}\mathrm{Cu}}\) collecting, respectively, Cu–O bonds with endpoints in \(\tilde D_{\mathrm{Cu}}\) and \(D_{\mathrm O}\), O–O bonds with both endpoints in \(D_{\mathrm O}\), and Cu–Cu bonds with both endpoints in \(\tilde D_{\mathrm{Cu}}\). The variational ansatz is then specified by the following mean-field Hamiltonian:
\begin{align}
H_{\text{mf}}^{\text{O-defect}} & =\sum_{\left\langle il\right\rangle ,\sigma}\chi_{il}\left(d_{i\sigma}^{\dagger}p_{l\sigma}+h.c.\right)-\sum_{\left\langle ll'\right\rangle ,\sigma}\chi_{ll'}\left(p_{l\sigma}^{\dagger}p_{l'\sigma}+h.c.\right)\nonumber \\
 & +\sum_{\left\langle ij\right\rangle ,\sigma}\chi_{ij}\left(d_{i\sigma}^{\dagger}d_{j\sigma}+h.c.\right)+\sum_{l,\sigma}\epsilon_{p}(l)p_{l\sigma}^{\dagger}p_{l\sigma}\nonumber \\
 & +\sum_{i,\sigma}\epsilon_{d}(i)d_{i\sigma}^{\dagger}d_{i\sigma}+\sum_{i}m_{Cu,i}(-1)^{x+y}\left(d_{i\uparrow}^{\dagger}d_{i\uparrow}-d_{i\downarrow}^{\dagger}d_{i\downarrow}\right)\nonumber \\
 & +\sum_{l}m_{O,l}(-1)^{x+y}\left(p_{l\uparrow}^{\dagger}p_{l\uparrow}-p_{l\downarrow}^{\dagger}p_{l\downarrow}\right),
\end{align}
where the hopping parameters are given by
\begin{align}
\chi_{il} & =\begin{cases}
\tilde{\chi}_{01} & \text{if}\;\left\langle il\right\rangle \in B_{\text{Cu-O}}\\
\chi_{01} & \text{otherwise},
\end{cases}\\
\chi_{ll'} & =\begin{cases}
\tilde{\chi}_{11} & \text{if}\;\left\langle ll'\right\rangle \in B_{\text{Cu-O}}\\
\chi_{11} & \text{otherwise},
\end{cases}\\
\chi_{ij} & =\begin{cases}
\tilde{\chi}_{00} & \text{if}\;\left\langle ij\right\rangle \in B_{\text{Cu-Cu}}\\
\chi_{00} & \text{otherwise},
\end{cases}
\end{align}
and those on-site terms are
\begin{align}
\epsilon_{d}(i),m_{\text{Cu},i} & =\begin{cases}
\epsilon_{d0},m_{z,00}^{(0)} & \text{if}\;i\in\tilde{D}_{\text{Cu}}\\
0,m_{z,00} & \text{otherwise}
\end{cases}\\
\epsilon_{p}(l),m_{\text{O},l} & =\begin{cases}
\epsilon_{p0},m_{z,11}^{(0)} & \text{if}\;l\in D_{\text{O}}\\
\epsilon_{p},m_{z,11} & \text{otherwise}.
\end{cases}
\end{align}
The correlation factors still take the form of Eq.~\eqref{eqn:vmc_gutz}.

\begin{figure}[t]
  \centering
  \includegraphics[width=0.48\textwidth]{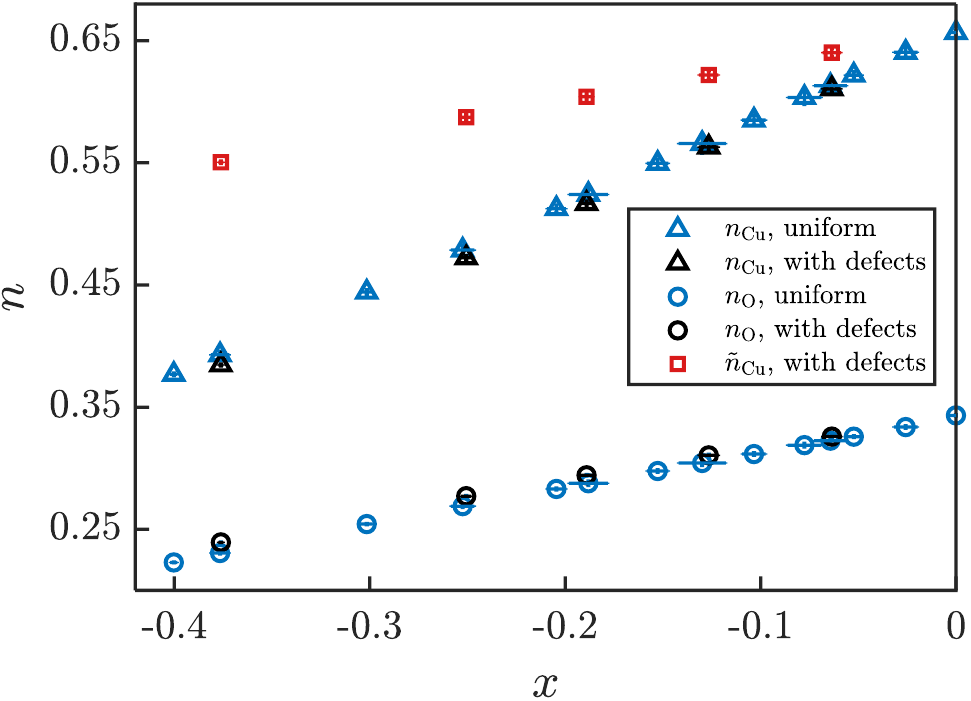}
  \caption{DQMC orbital-resolved hole densities. Triangles denote Cu sites; circles, O sites; red squares, the mean over nondefect Cu sites. Parameters match those in Fig.~\ref{fig8}(c). The DQMC charge distribution is consistent with the VMC results.}
  \label{fig_s4}
\end{figure}
\subsection{Supplemental numerical results}
In this subsection we present numerical results omitted from the main text. Fig.~\ref{fig_s4} shows orbital-resolved hole densities from DQMC for the Cu-defect (electron-doped) case, computed with the same parameters as in Fig.~\ref{fig8}(c) of the main text.

\end{document}